\documentclass{aa}
\usepackage{graphicx}
\usepackage{txfonts}
\usepackage{hyperref}
\usepackage{placeins}
\hypersetup{
    colorlinks=true,
    linkcolor=blue,
    filecolor=magenta,      
    urlcolor=blue,
    citecolor=blue
}

\newcommand{\Mc}{M_{\rm c}}

\newcommand{\Matm}{M_{\rm atm}}
\newcommand{\Mpl}{M_{\rm pl}}

\newcommand{\Miso}{M_{\rm iso}}

\newcommand{\Rgas}{\mathcal{R}}
\newcommand{\bigpar}[1]{\left( #1 \right)}
\newcommand{\alphav}{\alpha_{\rm v}}
\newcommand{\alphat}{\alpha_{\rm t}}
\newcommand{\AU}{{\rm AU}}
\newcommand{\bigbra}[1]{\left[ #1 \right]}
\begin{document}

\title{How stellar mass and disc size shape the formation\\and migration of super-Earths}
\author{Jesper Nielsen\inst{1,2}
\and
Anders Johansen \inst{1,2}
}

\institute{
Center for Star and Planet Formation, Globe Institute, University of Copenhagen, Øster
Voldgade 5-7, 1350 Copenhagen, Denmark email: \href{mailto:jesper.nielsen@sund.ku.dk}{jesper.nielsen@sund.ku.dk}
\and
Lund Observatory, Division of Astrophysics, Department of Physics, Lund University, Box 118, 22100 Lund, Sweden}

\date{Accepted 29 August 2025}

\abstract{The occurrence rate of close-in super-Earths is higher around M-dwarfs compared to stars of higher masses. In this work, we aim to understand how the super-Earth population is affected by the stellar mass, the size of the protoplanetary disc, and viscous heating. We utilised a standard protoplanetary disc model with both irradiated and viscous heating, together with a pebble accretion model, to simulate the formation and migration of planets. We find that if the disc is heated purely through stellar irradiation, inward migration of super-Earths is very efficient, resulting in the close-in super-Earth fraction increasing with increasing stellar mass. In contrast, when viscous heating is included, planets can undergo outward migration, delaying migration to the inner edge of the protoplanetary disc, which causes a fraction of super-Earth planets to grow into giant planets instead. This results in a significant reduction of inner super-Earths around high-mass stars and an increase in the number of giant planets, both of which mirror observed features of the planet population around high-mass stars. This effect is most pronounced when the protoplanetary disc is large, since such discs evolve over a longer timescale. We also tested a model when we injected protoplanets at a fixed time early on in the disc lifetime. In this case, we find that the fraction of close-in super-Earths decreases with increasing stellar mass in both the irradiated case and viscous case, since longer disc lifetimes around high-mass stars allow for planets to grow into giants instead of super-Earths for most injection locations.}

%\titlerunning{}
\authorrunning{Nielsen and Johansen}
\maketitle
\section{Introduction}
The observed exoplanet population provides extraordinary opportunities to constrain planet formation processes. With the detection of \textasciitilde 3000 exoplanets during its lifetime, the \textit{Kepler} telescope \citep{borucki2010_kepler} yielded many discoveries that have provided insights into the planet formation process. One such discovery is the apparent enhancement of the occurrence rate of close-in super-Earths for M-dwarfs compared to FGK-stars \citep{mulders2015_SEvMstar}. Most planets are thought to form in protoplanetary discs around young stars through core accretion, where the solid core is assembled first, followed by the accretion of a gaseous envelope. There are multiple mechanisms by which a protoplanet can accrete its solid core, such as through the accretion of planetesimals (kilometre-sized), pebbles (millimetre-sized), or a combination of the two \citep{alibert2018_hybridmodel,drazkowska2023_pfreview}\footnote{Given that pebble accretion can form both giants and terrestrial planets \citep{bitsch2015_giants,johansen2021_pebSS,onyett2023_pebSS}, we chose to include only pebble accretion in our model for simplicity.}. As the dust inventories of protoplanetary discs are observed to increase with increasing stellar mass \citep{pascucci2016_dustmasses,manara2023_pp7discs}, a first-order expectation would be that the occurrence rates of planets should also increase with increasing stellar masses. Indeed, this has been observed for giant planets \citep[e.g.][]{johnson2010_giantocc,reffert2015_GploccvMstar}. The mechanisms that led to a decrease of the super-Earth occurrence rate around high-mass stars are therefore currently uncertain.

As high-mass stars are more likely to form giant planets, it has been proposed that the early formation of an outer giant planet could block the influx of pebbles onto an inner planet, preventing it from growing to a super-Earth \citep[e.g.][]{mulders2021_MdwarfSE}. Such a model requires two important assumptions. The first is that the inner planet needs to form inside the water iceline and experience a reduction of the pebble flux due to the sublimation of water ice in the protoplanetary disc. This is required as otherwise the formation of the inner super-Earth is significantly faster than the formation of the outer giant due to the much higher efficiency of accreting drifting pebbles in the inner regions of protoplanetary discs \citep{danti2025_SEformGP}.\footnote{This assumption is not a strict requirement if the inner super-Earth starts growing significantly later than the outer giant.} The second assumption is that once the outer planet reaches pebble isolation mass, it blocks the entire pebble flux by creating a gap in the protoplanetary disc, quenching planet formation inside of its orbit. The degree to which giant planets are able to trap inward drifting pebbles for an extended period of time is a topic of intense current research \citep{haugbolle2019_filtering}. \citet{stammler2023_fluxblock} found that mm-sized pebbles are trapped by the planet and fragment into smaller sizes, which are sufficiently coupled to the gas to flow with the gas through the gap. Similar results were found by \citet{vanclepper2025_dustflowgap} and \citet{huang2025_filtering}, who performed 3D simulations around an embedded giant planet and found that small dust particles can indeed pass through the gap.
Further, outer giant planets have been found to have either a positive or zero correlation with the presence of inner super-Earths \citep{bryanlee2024_innerSE_outerGP,vanzandt2025_innerSE_outerGP,bryanlee2025_innerSPouterGP,bonomo2025_innerSEouterGP}. Should giant planets be able to block a significant amount of the pebble flux, it would be expected that outer giants are negatively correlated with the presence of inner super-Earths, in contrast with observations. Nevertheless, the presence of super-Jupiters has been found to be negatively correlated with the presence of an inner super-Earth, highlighting that the formation of very massive giant planets might indeed inhibit the formation of close-in super-Earths \citep{lefevremulders2025_innerSEoutersaturn}. Therefore, it is currently uncertain whether the lack of inner super-Earths around high-mass stars can be explained by the presence of an outer giant planet blocking the inward flux of pebbles.

The presence of binary companions around massive host stars has also been proposed as an explanation for the observed super-Earth population. Binary companions are more prevalent for massive stars and could suppress the formation of planets. However, even when accounting for the presence of binaries, \citet{morekratter2021_binariesSE} found that the enhanced occurrence rate of super-Earths around M dwarfs is only half of the observed enhancement.

Another proposed explanation is that, depending on the choice of model, the pebble accretion efficiency can be enhanced around low-mass stars \citep{chachan2023_SE_Mdwarf}. By utilising a planet population synthesis model, \citet{pan2025_SEvMstar} were recently able to explain the observed super-Earth occurrence rate as a function of stellar mass by a reduced migration rate around high-mass stars as well as through an increased pebble accretion efficiency around low-mass stars.

In this work, we aim to understand how the formation and migration of super-Earths are affected by the structure and sizes of protoplanetary discs. In Sect. \ref{sec:model}, we present our disc and planet formation model. In Sect. \ref{sec:results} we show the results of our planet population synthesis model, while in Sects. \ref{sec:discussion} and \ref{sec:conclusion}, we discuss and conclude our results, respectively. In Appendix \ref{app:photo_evap} we show the resulting fraction of close-in super-Earths for different disc photoevaporation models. In Appendix \ref{app:mig_coeff} we show the planet migration coefficient as a function of disc size and stellar mass as well as time, while in Appendix \ref{app:additional_figs} we show some additional simulation results using different model parameter values. 
\section{Planet formation model}
\label{sec:model}
\subsection{Protoplanetary disc model}
\label{ssec:ppd_model}
The viscosity of the disc, $\nu$, can be written as \citep{shakurasunyaev1973}
\begin{equation}
    \nu = \alphav c_{\rm s} H,
\end{equation}
where $\alphav$ is a dimensionless parameter setting the strength of the viscosity, $c_{\rm s}$ is the sound speed, and $H$ is the disc scale height. Throughout this work, we assumed a value of $\alphav = 0.01$. For a detailed discussion on our choice of $\alphav$, we refer to Sect. 7.1 of \citet{nielsen2023} and Sect. 2 of \citet{nielsen2025}. The sound speed can be written as 
\begin{equation}
    c_{\rm s} = \sqrt{\frac{k_{\rm b} T}{\mu_{\rm d}m_{\rm H}}} = \sqrt{\Rgas T},
\end{equation}
where $k_{\rm b}$ is the Boltzmann constant, $T$ is the temperature of the disc, $\mu_{\rm d}$ is the mean molecular weight of the disc, $m_{\rm H}$ is the mass of a hydrogen atom, and $\Rgas$ is the mass-specific gas constant. We use a value of 2.35 for $\mu_{\rm d}$, consistent with a gas that is composed of mostly H$_2$ and He. The gas scale height, in turn, can be written as
\begin{equation}
    H = \frac{c_{\rm s}}{\Omega},
\end{equation}
where $\Omega = \sqrt{GM_\star/r^3}$ is the Keplerian frequency, $M_\star$ is the stellar mass, $G$ is the gravitational constant, and $r$ is the distance between the planet and the star. The radial speed of the gas in the inner regions of the disc is
\begin{equation}
    u_r = -\frac{3}{2}\frac{\nu}{r}.
\end{equation}
We used two different models to calculate the temperature in the midplane of the disc. The first is a purely irradiated disc, where the temperature is given as \citep{ida2016}
\begin{equation}
    T_{\rm irr} = 150\,{\rm K}\bigpar{\frac{M_\star}{M_\odot}}^{-1/7}\bigpar{\frac{L_\star}{L_\odot}}^{2/7}\bigpar{\frac{r}{\AU}}^{-3/7}.
\end{equation}
We scaled the stellar luminosity, $L_\star$, as
\begin{equation}
\label{eq:L_star}
    \frac{L_\star}{L_\odot} = \bigpar{\frac{M_\star}{M_\odot}}^2,
\end{equation}
which is typical for young stars \citep{baraffe2002_lum,baraffe2015_lum}.  

In our second model, we also included viscous heating\footnote{Sometimes also referred to as accretion heating.} of the disc, which we took from \citet{ida2016} as
\begin{equation}
    T_{\rm visc} = 317\,{\rm K}\bigpar{\frac{M_\star}{M_\odot}}^{3/10}\bigpar{\frac{\alphav}{0.01}}^{-1/5}\bigpar{\frac{\dot{M}_{\rm g}}{10^{-7}\,{\rm M_\odot/yr}}}^{2/5}\bigpar{\frac{r}{1\,\AU}}^{-9/10},
\end{equation}
where $\dot{M}_{\rm g}$ is the gas accretion rate onto the star. When we include viscous heating, the temperature of the disc is given as max($T_{\rm irr}$,$T_{\rm visc}$).  

From the continuity equation, the surface density of the gas in the accreting regions of the disc, $\Sigma_{\rm g}$, relates to $\dot{M}_{\rm g}$ as
\begin{equation}
    \Sigma_{\rm g} = -\frac{\dot{M}_{\rm g}}{2\pi r u_r} = \frac{\dot{M}_{\rm g}\sqrt{GM_\star}}{3\pi\alphav\Rgas T r^{3/2}}.
\end{equation}
For the irradiated case, this gives a surface density of
\begin{equation}
\begin{split}
    \Sigma_{\rm g,irr} & \approx 2500\,{\rm g/cm^2} \bigpar{\frac{L_\star}{L_\odot}}^{-2/7}\bigpar{\frac{M_\star}{M_\odot}}^{9/14}\bigpar{\frac{\alphav}{0.01}}^{-1} \\ 
    & \times \bigpar{\frac{\mu_{\rm d}}{2.35}}\bigpar{\frac{\dot{M}_{\rm g}}{10^{-7}\,{\rm M_\odot/yr}}}\bigpar{\frac{r}{1\, \AU}}^{-15/14},
\end{split}
\end{equation}
while in the viscously heated regions, the surface density is given by
\begin{equation}
\begin{split}
    \Sigma_{\rm g,visc} & \approx 1325\,{\rm g/cm^2}\bigpar{\frac{M_\star}{M_\odot}}^{1/5}\bigpar{\frac{\alphav}{0.01}}^{-4/5} \\
    & \times \bigpar{\frac{\dot{M}_{\rm g}}{10^{-7}\,{\rm M_\odot/yr}}}^{3/5}\bigpar{\frac{r}{1\, \AU}}^{-3/5}.
\end{split}
\end{equation}
When viscous heating is taken into account, the surface density is given by min($\Sigma_{\rm g,irr}$,$\Sigma_{\rm g, visc}$).

The time evolution of the disc accretion rate can be written as \citep{hartmann1998}
\begin{equation}
    \dot{M}_{\rm g} = \dot{M}_{\rm g,0}\left(\frac{t}{t_{\rm s}}+1\right)^{((5/2)-\gamma)/(2-\gamma)},
\end{equation}
where $\gamma$ is the power-law index of the viscosity, as a function of distance to the star, $r$. The characteristic evolution timescale, $t_{\rm s}$, is given by 
\begin{equation}
\label{eq:t_s}
    t_{\rm s} = \frac{1}{3(2-\gamma)^2}\frac{R_{\rm d,0}^2}{\nu(R_{\rm d,0})},
\end{equation}
where $R_{\rm d,0}$ is the initial disc size. The initial gas accretion rate, $\dot{M}_{\rm g,0}$, at $t=0$, is a free parameter. The estimated initial sizes of protoplanetary discs are still quite uncertain and very model dependent. Using observations of CO, gas disc sizes have been estimated to be of the order of $\sim$100 AU \citep{ansdell2018_dustsizes}, although recent work by \citet{tabone2025_accmech} suggests that the initial disc sizes are much smaller at $\sim$10 AU. Nevertheless, it is an important parameter, as in typical protoplanetary disc models the disc size determines the characteristic timescale for the evolution of the gas in the disc. Further, for discs with a given mass, a smaller disc would indicate higher accretion rates and therefore higher surface densities, which in turn affects the formation and migration of planets \citep[e.g.][]{drazkowska2021_pebbleflux}. We therefore treat the disc size as a free parameter in this work. Throughout this work, we set the initial accretion rates, $\dot{M}_{\rm g,0}$, according to the observed power law from \citet{almendros-abad2024_discrelation}, who found that the accretion rates scale as
\begin{equation}
\label{eq:Mdot_Mstar}
    \dot{M}_{\rm g} \propto M_\star^{1.67}
\end{equation}
for the Ophiuchus star-forming region, which was the youngest region they considered. We normalised the accretion rate such that a solar-mass star has an initial accretion rate of $10^{-7}$ $M_\odot$/yr, which is typical for young solar-mass stars \citep{hartmann2016}.

The power-law index of the viscosity, $\gamma$, was determined by the power-law index of the temperature. In the model when we included viscous heating, we used the power-law index for the irradiated case since the length scale for the disc evolution is the initial disc size, which is typically much larger than the transition radius, $r_{\rm tran}$, where the disc becomes viscously heated \citep{liu2019_miso}. The initial mass of the disc in the irradiated case is given by 
\begin{equation}
\label{eq:M_gas_irr}
\begin{split}
    M_{\rm d,irr,0} & = \int_0^{R_{\rm d,0}} 2\pi r \Sigma_{\rm g} dr  = 0.1\,M_\odot\bigpar{\frac{R_{\rm d,0}}{71\, \AU}}^{13/14} \bigpar{\frac{\mu_{\rm d}}{2.35}} \\
    & \times \bigpar{\frac{\alphav} {0.01}}^{-1}\bigpar{\frac{\dot{M}_{\rm g,0}}{10^{-7}\,M_\odot/{\rm yr}}}\bigpar{\frac{L_\star}{L_\odot}}^{-2/7}\bigpar{\frac{M_\star}{M_\odot}}^{9/14},
\end{split}
\end{equation}
In the case of pure viscous heating, the initial disc mass is given by
\begin{equation}
\label{eq:M_gas_visc}
\begin{split}
    M_{\rm d,visc} & = 0.26\,M_{\odot}\bigpar{\frac{M_\star}{M_\odot}}^{1/5}\bigpar{\frac{\dot{M}_{\rm g,0}}{10^{-7}\,{\rm M_\odot/yr}}}^{3/5} \\
    & \times \bigpar{\frac{\alphav}{0.01}}^{-4/5}\bigpar{\frac{R_{\rm d,0}}{71\,\AU}}^{7/5}.
\end{split}
\end{equation}
The initial disc mass in the viscously heated case is therefore a combination of Eqs. \eqref{eq:M_gas_irr} and \eqref{eq:M_gas_visc}
\begin{equation}
\label{eq:M_gas_full_viscous}
\begin{split}
    M_{\rm d,full} & = \int_0^{R_{\rm d,0}} 2\pi r \Sigma_{\rm g} dr = \int_0^{r_{\rm tran}}2\pi r \Sigma_{\rm visc} dr +\int_{r_{\rm tran}}^{\rm R_{\rm d,0}}2\pi r\Sigma_{\rm irr}dr \\
    & = M_{\rm d,visc}(r_{\rm tran}) + M_{\rm d,irr}-M_{\rm d,irr}(r_{\rm tran}),
\end{split}
\end{equation}
where $M_{\rm d,visc}(r_{\rm tran})$ is the mass enclosed at $r_{\rm tran}$ in the fully viscous case and $M_{\rm d,irr}(r_{\rm tran})$ is the same but for the irradiated case. The transition radius between the viscously heated region and the irradiated region is given by setting $T_{\rm irr} = T_{\rm visc}$, and yields
\begin{equation}
\begin{split}
    r_{\rm tran}  & \approx 4.9\,\AU\bigpar{\frac{M_\star}{M_\odot}}^{31/33}\bigpar{\frac{\alphav}{0.01}}^{-14/33} \\
    & \times \bigpar{\frac{\dot{M}_{\rm g}}{10^{-7}\,M_\odot/{\rm yr}}}^{28/33}\bigpar{\frac{L_\star}{L_\odot}}^{-20/33}.
\end{split}
\end{equation}
As mentioned, $r_{\rm tran,0}$$\,\ll\,$$ R_{\rm d,0}$, which means that $M_{\rm d,full,0}$$\,\approx\,$$M_{\rm d,irr,0}$ so for simplicity, we set the disc mass to be $M_{\rm d,irr,0}$.

We also took X-ray photoevaporation of the gas disc into account by following the mass loss relation from \citet{owen2012_photoevap}
\begin{equation}
    \dot{M}_{\rm w} = 6.25\cdot10^{-9}\,{\rm M_\odot/yr}\bigpar{\frac{M_\star}{M_\odot}}^{-0.068}\bigpar{\frac{L_{\rm X}}{10^{30}\,{\rm erg/s}}}^{1.14},
\end{equation}
where the X-ray luminosity, $L_{\rm X}$, is given by \citet{bae2013_xray} as
\begin{equation}
    \log(L_{\rm X}[{\rm erg/s}]) = 30.37+1.44\bigpar{\frac{M_\star}{M_\odot}}.
\end{equation}
When the photoevaporation rate is larger than the accretion rate onto the star, photoevaporation dominates. When this occurred, we followed \citet{liu2019_miso} and modified the accretion rate to
\begin{equation}
     \dot{M}_{\rm g} = \dot{M}_{\rm g,0}\left(\frac{t}{t_{\rm s}}+1\right)^{((5/2)-\gamma)/(2-\gamma)}\exp\bigpar{-\frac{t-t_{\rm pho}}{\tau_{\rm pho}}},
\end{equation}
where $t_{\rm pho}$ is the time $t$ where $\dot{M}_{\rm g} = \dot{M}_{\rm W}$. The mass-loss timescale, $\tau_{\rm pho}$, is given by
\begin{equation}
    \tau_{\rm pho} = \frac{M_{\rm d}(t=t_{\rm pho})}{\dot{M}_{\rm W}}.
\end{equation}
We then simply set the disc lifetime to be $t_{\rm pho}+\tau_{\rm pho}$. For solar-mass stars with an initial accretion rate of $10^{-7}\,{\rm M_\odot/yr}$, and an initial disc mass of 0.1 $M_\odot$, this results in a lifetime of 4.8 Myr. We also tested the updated photoevaporation model presented in \citet{ercolano2023_discPE}, which calculates how the mass-loss rate depends on stellar mass and X-ray luminosity based on the works by \citet{picogna2021_discPE} and \citet{ercolano2021_PELx}, respectively. That model predicts higher values for $\dot{M}_{\rm w}$ and thus lower disc lifetimes for different stellar masses, with the disc lifetime around a solar-mass star being $\sim$2 Myr. Nevertheless, we found little difference in the planet population between the two photoevaporation models, as the accretion rate and consequently gas surface density decreases significantly at these later stages of the protoplanetary disc. This means that extending the disc lifetime with a few megayears does not have a major effect on the overall planet population. We show these results in Appendix \ref{app:photo_evap}.

\begin{figure}
    \centering
    \includegraphics[width=\linewidth]{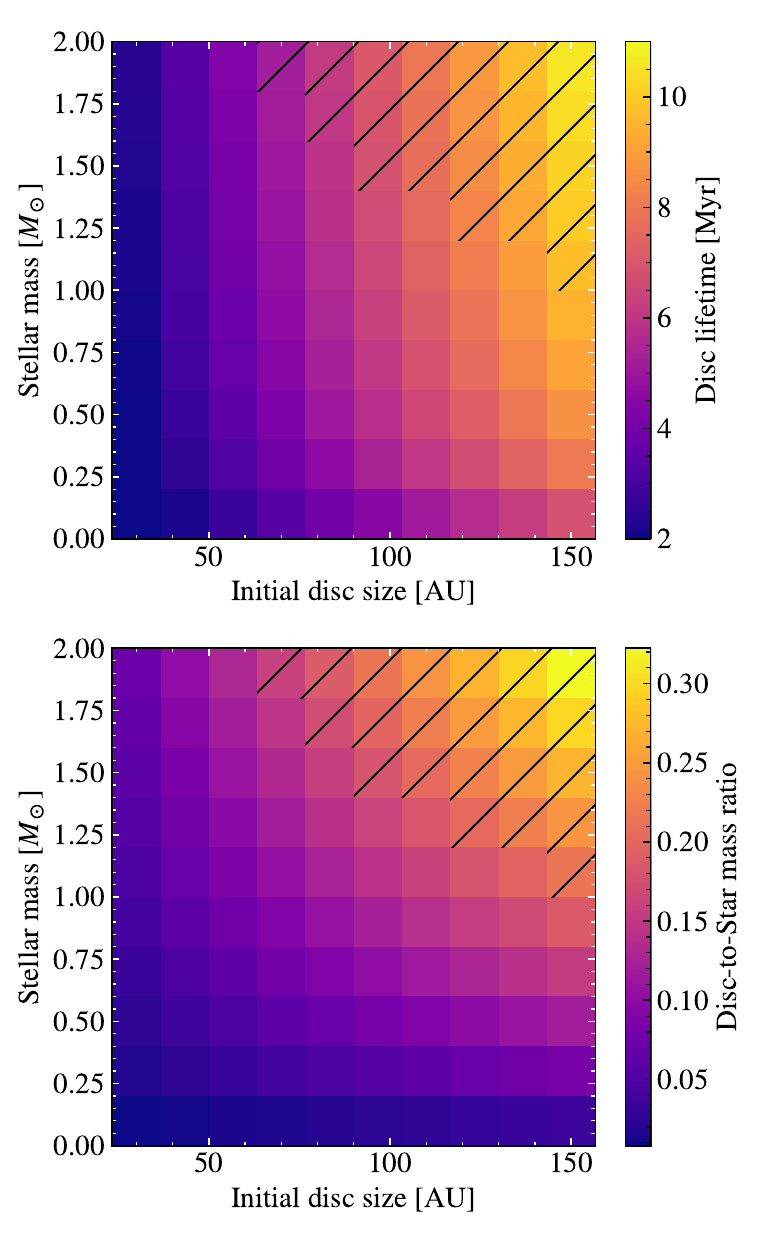}
    \caption{Disc lifetimes (top) and disc-to-star mass ratios (bottom) for discs with varying stellar masses and initial disc sizes. For a fixed disc mass, the disc lifetime increases with increasing stellar mass, while a larger disc also results in longer disc lifetime. Large discs ($\gtrsim$100 AU) around stars more massive than $\sim$1.5 $M_\odot$ are gravitationally unstable in some regions of the discs. This region is marked by a hatched pattern.}
    \label{fig:lifetimes}
\end{figure}

We show disc lifetimes as well as the disc-to-star mass ratios for a range of stellar masses and disc sizes in Fig. \ref{fig:lifetimes}. Large discs around massive host stars are longer-lived. However, discs larger than 100 AU around stars more massive than $\sim$1.5 $M_\odot$ are typically gravitationally unstable in their outer regions (marked by a hatched pattern). Further, even if the disc lifetime is $\sim$10 Myr, the disc is pretty inactive in forming planets after $\sim$3 Myr due to the low gas surface densities at these late times. The ratio between disc masses and stellar masses ranges from $\sim$1\% to $\sim$30\%.

From this protoplanetary disc model, it is clear that the disc size plays an important role in setting important physical properties of the disc. For a given accretion rate, the disc size determines the characteristic timescale of the disc, as seen in Eq. \eqref{eq:t_s}. Further, the disc size also determines the mass of the disc as seen in Eq. \eqref{eq:M_gas_irr}. The mass of the gas disc also sets the mass budget of solids in the disc, which in turn affects which planets can form. Therefore, it might be more intuitive to have the disc mass as a free parameter instead of the disc size. However, as the disc size also directly affects the disc accretion rate through $t_{\rm s}$, we argue that the disc size is the more natural parameter to vary. Moreover, disc sizes can to some extent be observed directly as opposed to disc masses, which have to be inferred indirectly \citep[see e.g.][and references therein]{manara2023_pp7discs}.

Finally, we modelled the inner edge of the disc following \citet{mulders2015_SEvMstar}, where the inner edge is approximated to the co-rotation radius of the star given by the expression
\begin{equation}
\label{eq:r_in}
    r_{\rm in} = 0.04\bigpar{\frac{M_\star}{M_\odot}}^{1/3}.
\end{equation}

\subsection{Core accretion}
\label{ssec:core_acc} 
Our pebble accretion model is based on the work of \citet{drazkowska2021_pebbleflux} and is detailed in \citet{nielsen2023}. Throughout this work, we assumed that the star has a solar composition, with the elemental composition based on the work by \citet{magg2022_solarcomp}. The resulting dust-to-gas ratio in the disc is therefore 0.014. We calculated the growth of dust in the protoplanetary disc due to sticking by micrometre-sized monomers. The dust size is limited by fragmentation (both turbulent fragmentation and fragmentation from differential drift) as well as radial drift. We set the fragmentation velocity of pebbles to 2 m/s throughout the entire disc so that the Stokes numbers, St, of pebbles in the outer regions are $\sim$0.01. Typical reported fragmentation velocities range between 1 m/s for silicate particles and 10 m/s for icy particles \citep[e.g.][]{gundlach_blum2015_vfrag}. However, as shown by \citet{nielsen2025} in their appendix A, varying the fragmentation velocity based on the composition of the dust does not have a significant effect on the final super-Earth population. The core then grows following Eqs. (2)-(5) and (20)-(28) in \citet{nielsen2023}.
\\\\
Pebble accretion continues until the pebble isolation mass, $\Miso$, is reached. We calculated the pebble isolation mass following \citet{bitsch2018_Miso},
\begin{equation}
\begin{split}
        \Miso  & = 25 M_\oplus \bigpar{\frac{H/r}{0.05}}^3\bigbra{0.34\bigpar{\frac{-3}{\log(\alpha_\mathrm{t})}}^4+0.66} \\
        & \times \bigpar{1+\frac{\chi-2.5}{6}}\bigpar{\frac{M_*}{M_\odot}},
\end{split}
\end{equation}
where $\alphat$ is a parameter that defines the turbulence strength and is set in this work to $10^{-4}$. This $\alpha$-parameter is distinct from the $\alpha$-parameter that drives the loss of angular momentum in the disc and subsequent gas flux onto the star, $\alphav$, as it has been shown that these two parameters need not be equal \citep{lesur2023_pp7discreview}. The parameter $\chi$ is the negative logarithmic pressure gradient of the gas. In the irradiated regime, this is equal to $39/14\approx2.786$ while in the viscous regime, it is equal to $51/20=2.55$.

\subsection{Planet migration}
\label{ssec:migration}
\begin{figure}
    \centering
    \includegraphics[width=\linewidth]{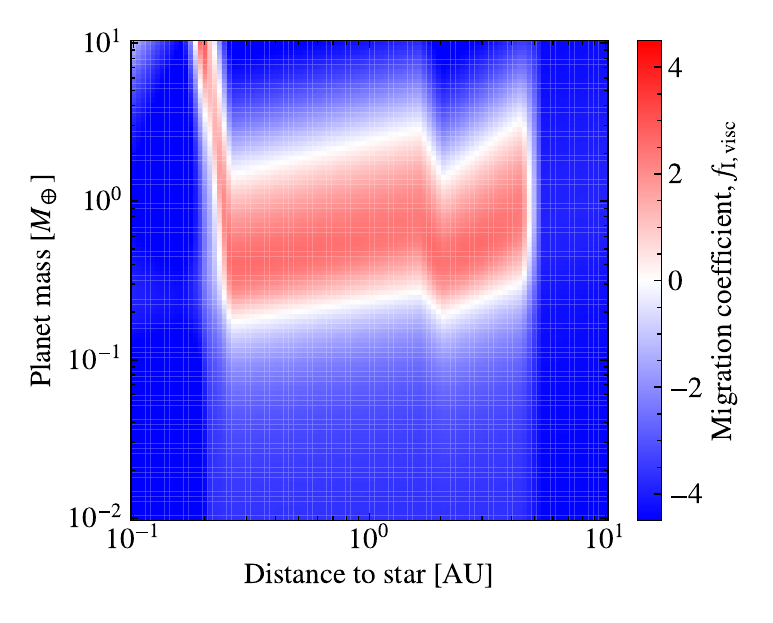}
    \caption{Migration coefficient, $f_{\rm I,visc}$, as a function of distance to the star and planet mass for a solar-mass star with a disc size of 71 AU at $t=0$. Inside of the transition radius ($\sim$5 AU), the disc is viscously heated, and for certain masses, the planet experiences outward migration. The sharp switch to inward migration at $\sim$0.2 AU is caused by the significant decrease in opacity due to the evaporation of silicate and metal grains, which occurs at $\sim$1000 K. We also show $f_{\rm I, visc}$ at different times and for different stellar masses and disc sizes in appendix \ref{app:mig_coeff}.}
    \label{fig:mig_coeff}
\end{figure}

The migration of the protoplanet is driven by the excitation of density wakes as a result of gravitational planet-disc interactions \citep{kleynelson2012_migration}. While it has been shown that the dust in the disc can also induce torques onto the migrating planet, which can cause halted inward migration as well as outward migration if the dust-to-gas ratios in the disc are high enough \citep{benitez2018_dusttorques,guilera2025_dusttorques}. We chose to neglect these effects for simplicity. Neglecting these effects, type-I migration for a fully irradiated disc can be written as 
\begin{equation}
\label{eq:rdot_irr}
    \dot{r}_{\rm I,irr} = f_{\rm I,irr}\frac{\Mpl}{M_\star}\frac{\Sigma_{\rm g}r^2}{M_\star}\bigpar{\frac{H}{r}}^{-2}v_{\rm K},
\end{equation}
where $v_{\rm K}$ is the Keplerian speed at the location of the planet. The factor, $f_{\rm I,irr}$, is a constant that has been fit by \citet{dangelo_lubow2010} to be
\begin{equation}
\label{eq:fmig_irr}
    f_{\rm I,irr} = -2(1.36+0.62\frac{\partial \ln\Sigma_{\rm g}}{\partial \ln r}+0.43\frac{\partial \ln T}{\partial \ln r}) \approx -4.42.
\end{equation}
However, when the disc is viscously heated, a more complex migration law needs to be taken into account. Therefore, when we include viscous heating in our model, we follow \citet{paardekooper2011_migration}, specifically their Eqs. (50)-(53), and calculate the total torque, $\Gamma$, onto the planet as the sum of the Lindblad torque, $\Gamma_{\rm L}$ and the co-rotation torque, $\Gamma_{\rm co}$. The migration rate can then be written as
\begin{equation}
    \dot{r}_{\rm I} = \frac{2\Gamma}{\Mpl v_{\rm K}}.
\end{equation}
This can be rewritten to be in the form of Eq. \eqref{eq:rdot_irr}\footnote{Eqs. (50)-(53) in \citet{paardekooper2011_migration} are also valid for an irradiated disc and yield a similar migration coefficient as the one found by \citet{dangelo_lubow2010}. Therefore, we used the relation from \citet{dangelo_lubow2010} to calculate the migration in the irradiated case due to it being computationally faster.} but with $f_{\rm I,irr}$ replaced with $f_{\rm I, visc}$, which, similarly to $f_{\rm I,irr}$, depends on the surface density- and temperature gradients, but also on $\alphat$ and the opacity of the disc. For the opacity of the disc, which enters the migration rate in the viscously heated case, we use the opacity law from \citet{bell_lin1994_opacities}. In Fig. \ref{fig:mig_coeff}, we show the value of the migration coefficient, $f_{\rm I,visc}$, as a function of planet mass and distance to the star for a solar-mass star with a disc size of 71 AU at $t=0$. Clearly, there is a large region of outward migration within $r_{\rm tran}$, which prevents super-Earths from migrating inwards significantly. This can be contrasted to discs heated purely through irradiation where type-I migration is stronger than in the viscous case and always in the direction towards the star, as $f_{\rm I,irr}$ is in that case negative with a value of -4.42 compared to $f_{\rm I,visc}$, which varies between -4 and 2 depending on the planet mass and distance to the star. Further, inside of $r_{\rm tran}$, the surface density of the star is by definition lower in the viscous case compared to the purely irradiated case, which means that migration is even slower in this region in the viscous case compared to the irradiated case.
\\\\
Should the planet grow large enough, the planet can start to carve out a gap in the disc \citep{kanagawa2018,paardekooper2023_pp7migreview}. This results in a transition towards type-II migration, where the depth of the gap in the disc affects the rate at which the planet migrates inwards. We included type-II migration by modifying our migration equation following \citet{johansen2019},
\begin{equation}
\label{eq:mig_eq}
    \dot{r} = \dot{r}_{\rm I}\frac{1}{1+[\Mpl/M_{\rm gap}]^2},
\end{equation}
where $M_{\rm gap}$ is the mass at which the planet starts to carve a gap into the surrounding gas, which we set to be equal to 2.3 times $\Miso$ following the approach of \citet{johansen2019}. This definition of the gap transition mass ensures that it never exceeds the pebble isolation mass, which otherwise might be possible at the low levels of turbulence that we assume in this work. Additionally, how the pebble isolation mass scales with $\alphat$ is currently uncertain, see section 3.1 in \citet{johansen2019} for more detailed discussions. As such, we chose a more simplified definition of the gap transition mass. Finally, we allowed planets to grow and migrate until they reached the inner edge, $r_{\rm in}$, of the disc, after which migration and growth was halted.

\subsection{Gas accretion}
\label{ssec:gas_acc}
After the pebble isolation mass was reached, the planet began to cool and accrete gas. We followed \citet{bitsch2015b_gasaccretion}, who derived an analytical model based on the work of \citet{piso_youdin2014}, to yield
\begin{equation}
\begin{split}
    \dot{M}_{\rm atm, KH} & = 1.75\times 10^{-4} M_\oplus \,{\rm yr}^{-1} f^{-2} \bigpar{\frac{\kappa_{\rm env}}{1\,{\rm cm^2\, g^{-1}}}}^{-1} \\
    & \times \bigpar{\frac{\rho_{\rm c}}{5.5\, {\rm g\, cm^{-3}}}}^{-1/6}\bigpar{\frac{\Mc}{M_\oplus}}^{11/3}\bigpar{\frac{\Matm}{M_\oplus}}^{-1}\bigpar{\frac{T_{\rm d}}{81\, {\rm K}}}^{-1/2},
\end{split}
\end{equation}
where $\kappa_{\rm env}$ is the envelope opacity, which we took to be 0.05 cm$^2$ g$^{-1}$. The numerical factor, $f$, is included to match the analytical results to the numerical and is set to 0.2. This analytical gas accretion rate is only valid when the planet is not undergoing runaway gas accretion, i.e. when $\Matm < \Mc$. In the case of $\Matm \geq \Mc$, we followed \citet{machida2010_gasacc} who estimated the rapid gas accretion rate from 3D hydrodynamical simulations
\begin{equation}
    \dot{M}_{\rm atm,rapid} = 
    \begin{cases}
        0.83\Omega\Sigma_{\rm g}H^2\bigpar{\frac{R_{\rm H}}{H}}^{9/2} & \quad {\rm if} \quad \frac{R_{\rm H}}{H} < 0.3 \\
        0.14\Omega\Sigma_{\rm g}H^2 & \quad {\rm if} \quad \frac{R_{\rm H}}{H} \geq 0.3.
    \end{cases}
\end{equation}
These accretion rates are limited by the amount of gas that can be supplied to the planet from the protoplanetary disc. \citet{ida2018} found that the gas can enter the Hill sphere at a rate of
\begin{equation}
\begin{split}
    \dot{M}_{\rm atm,hill}  & = 1.5\times 10^{-3}\,M_\oplus\,\mathrm{yr^{-1}}\bigpar{\frac{H/r}{0.05}}^4 \bigpar{\frac{\Mpl}{10M_\oplus}}^{4/3} \\ 
    & \times \bigpar{\frac{\alphav}{0.01}}^{-1}\bigpar{\frac{\dot{M}_\mathrm{g}}{10^{-8}M_\odot\mathrm{yr^{-1}}}}\frac{1}{1+(M/M_\mathrm{gap})^2}.
\end{split}
\end{equation}
Finally, the gas accretion rate cannot exceed the gas flux that flows through the disc, $\dot{M_{\rm g}}$. We therefore set the gas accretion rate to be
\begin{equation}
    \dot{M}_{\rm atm} = 
    \begin{cases}
        {\rm min}(\dot{M}_{\rm atm, KH}, \dot{M}_{\rm atm, Hill}, \dot{M}_{\rm g}) & \quad {\rm if} \quad \Matm < \Mc \\ 
        {\rm min}(\dot{M}_{\rm atm, rapid}, \dot{M}_{\rm atm, Hill}, \dot{M}_{\rm g}) & \quad {\rm if} \quad \Matm \geq \Mc.
    \end{cases}
\end{equation}
\section{Results}
\label{sec:results}
\begin{figure}
    \centering
    \includegraphics[width=\linewidth]{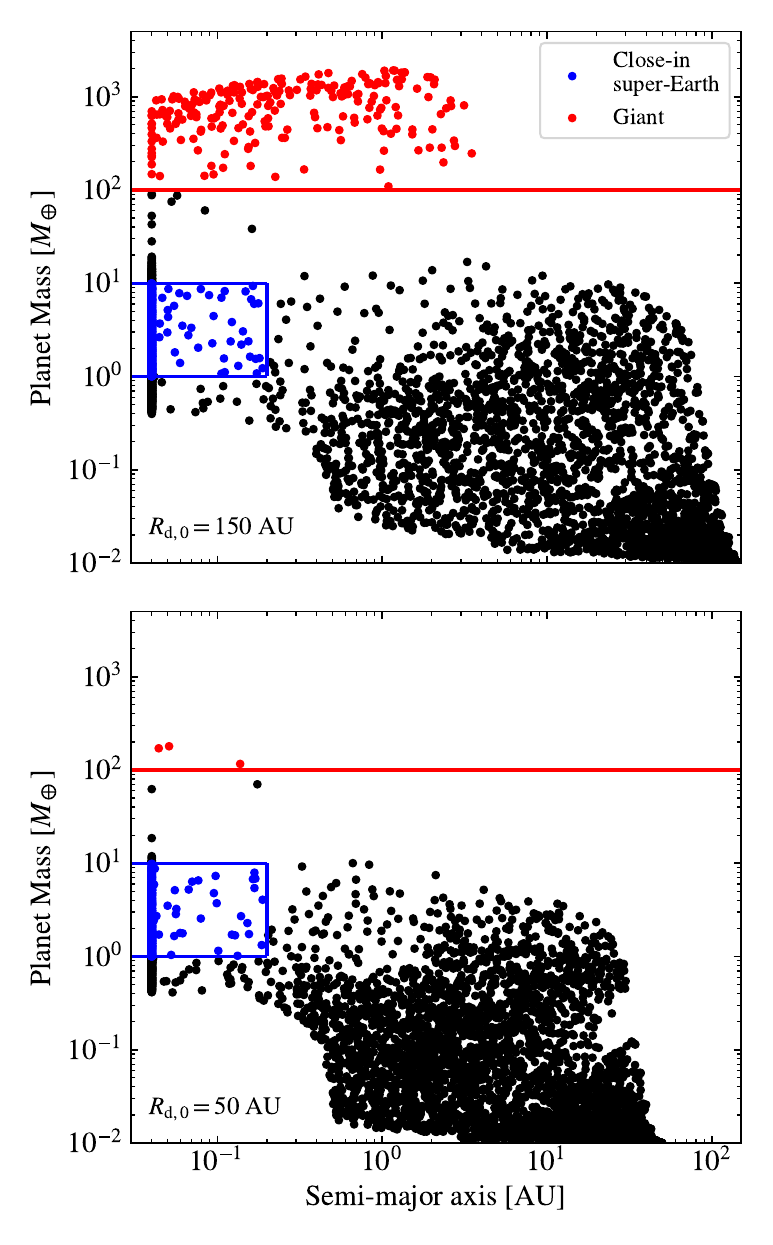}
    \caption{Example of the resulting semi-major axes and masses of planets injected around a solar-mass star for two different initial disc sizes, 150 AU (top) and 50 AU (bottom). Here, only irradiation heating is considered. We indicate the planets that we define as close-in super-Earths (blue) and giants (red). The large disc forms significantly more giant planets due to having more available mass and a longer lifetime.}
    \label{fig:sun_categories}
\end{figure}

We injected our planets by sampling the beginning location from a log-uniform distribution between 0.5 AU and the initial disc size, $R_{\rm d,0}$. The injection time was sampled from a uniform distribution between $10^3$ yr and the disc lifetime, $t_{\rm d}$. The initial protoplanet mass was set to be 0.01 $M_\oplus$, which allowed us to ignore Bondi accretion and only work within the 2D and 3D Hill regimes of pebble accretion \citep{drazkowska2023_pfreview}. We thereby assumed that embryos form rapidly through pebble accretion \citep{lyra2023_growth} or through a combination of planetesimal accretion and pebble accretion \citep{lorek_johansen2022_planetesimalacc}. In all simulations throughout this work, we injected 5000 protoplanets. In other words, for each protoplanetary disc, we sampled 5000 combinations of injection locations and injection times. We defined super-Earths as those with masses between 1 and 10 $M_\oplus$ to facilitate comparison with previous theoretical work \citep[e.g.][]{pan2024_collisions}. We also explored different definitions of close-in planets. The upper semi-major axis limit for planets around M-dwarfs in \citet{mulders2015_SEvMstar} is 0.2 AU. In comparison, \citet{pan2025_SEvMstar} defined close-in planets as planets with orbital periods of less than 100 days. We compared our results using both of these definitions and found no significant difference between them. Therefore, we chose to define close-in planets as those planets with semi-major axes less than 0.2 AU. We show the resulting semi-major axes and masses of 5000 planets injected in two example discs with sizes of 50 AU and 150 AU, respectively, around solar-mass stars in Fig. \ref{fig:sun_categories}. Here, the disc is only heated through stellar irradiation. We also show our definitions of close-in super-Earths and giants in blue and red, respectively. Even with rapid inwards type-I and type-II migration in the disc, we can form Jupiter-mass planets as far out as 4 AU if the protoplanetary disc is large enough. The formation of giants on wider orbits can also be facilitated if the pebble flux in the outer regions decays significantly, which allows the planet to cool sufficiently and thus starts accreting gas \citep{gurrutxaga2023_outergiant}; however, we ignore that possibility here.

\subsection{Exploring the parameter space}
\label{ssec:parameter_space}
\begin{figure*}
    \centering
    \includegraphics[width=\linewidth]{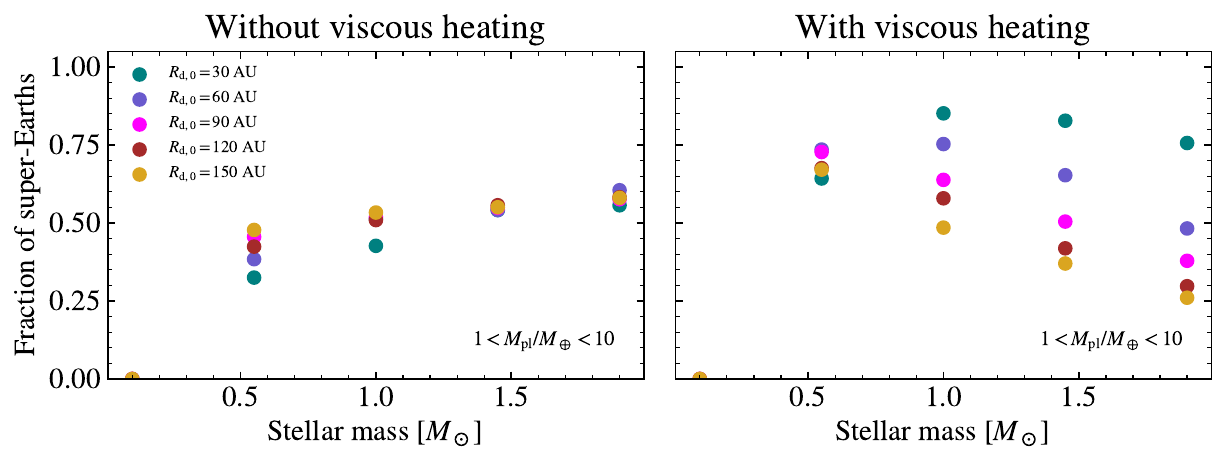}
    \caption{The fraction of close-in super Earths ($r < 0.2$ AU) for a range of stellar masses (x-axis) and disc sizes (colours). In all cases, no super-Earths are formed around stars with a mass of $0.1\, M_\odot$. When only considering discs heated by irradiation (left), the fraction of super-Earths increases with increasing stellar mass regardless of disc size. In the viscously heated case (right), the fraction of super-Earths reaches a maximum at 0.5 $M_\odot$ and then declines slightly for large discs (90 AU or larger). For smaller discs (30 and 60 AU), the super-Earth fraction peaks at 1 $M_\odot$, while for larger discs it peaks at 0.5 $M_\odot$.}
    \label{fig:scaling_disc_SE_close_frac}
\end{figure*}

In this section, we explore the parameter space of disc sizes and stellar masses. For a given disc size and initial accretion rate, we calculated the gas mass of the disc from Eq. \eqref{eq:M_gas_irr}. In Fig. \ref{fig:scaling_disc_SE_close_frac}, we show the resulting close-in super-Earth fraction\footnote{We define the close-in super-Earth fraction to be the number of close-in super-Earths, $N_{\rm close, SE}$, divided by the number of close planets, $N_{\rm close}$, where a close planet is defined as a planet with $r<0.2$ AU. We discuss this definition and its application to observations further in Sect. \ref{ssec:SE_frac_discussion}.} as a function of stellar mass and disc size.

In the irradiated case, the fraction of close-in super-Earths increases with increasing stellar mass for all disc sizes. For a given stellar mass, disc size has a minor effect on the super-Earth fraction as even small discs form super-Earths. Clearly, the irradiated case is unable to reproduce the observed super-Earth trend, which peaks at $\sim$$0.5$ $M_\odot$. When considering viscous heating, however, we find a marked decrease with increasing stellar mass in the resulting super-Earth fraction for nearly all disc sizes. For the largest discs (90-150 AU), the super-Earth fraction peaks at 0.5 $M_\odot$ while for smaller discs, the peak is shifted towards solar-mass stars. 

\begin{figure*}
    \centering
    \includegraphics[width =\linewidth]{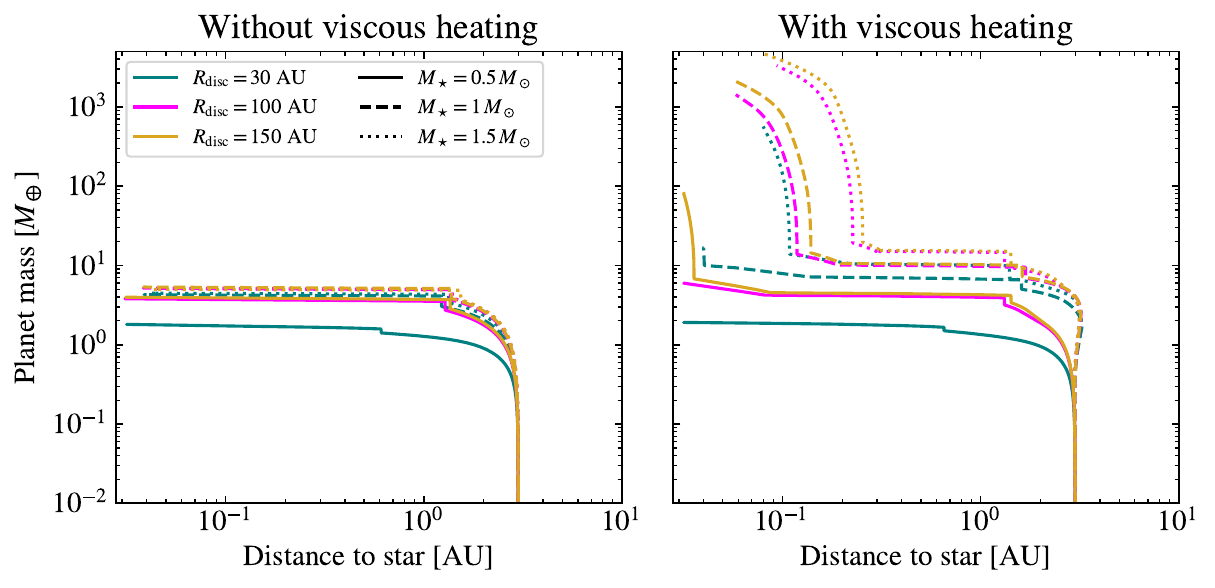}
    \caption{Growth tracks of a single planet injected at $r=3$ AU and $t=10^4$ yr. In the irradiated case (left), this results in the formation of a super-Earth at the inner disc edge for all stellar masses and disc sizes. In the viscous heating case (right), the protoplanet grows into a giant planet for high stellar masses as well as for solar-mass stars with sufficiently large discs.}
    \label{fig:alpha_scaling_oneplanet}
\end{figure*}

In order to understand the cause of the reduction of the close-in super-Earth fraction around high-mass stars, we investigated some example growth tracks of individual planets. We injected the protoplanet at $r=3$ AU and $t = 10^4$ yr around stars with a range of disc sizes and masses and show the growth tracks in Fig. \ref{fig:alpha_scaling_oneplanet}. When the disc is heated solely through irradiation, inward migration is very efficient and the planet grows into a super-Earth with a final orbit at the inner edge for all stellar masses and disc sizes. In the viscous case, similar results are found for disc sizes of 30 and 100 AU around a star with a mass of 0.5 $M_\odot$. In the case of a disc with a size of 150 AU and $M_\star = 0.5\, M_\odot$, the planet has just enough time to grow into a Saturn-mass planet. Around a solar-mass star, only a small disc size yields a super-Earth as now the planet experiences outward migration, delaying the migration to the inner edge, which allows the planet core to become massive enough for the planet to grow into a giant. For a star with a mass of 1.5 $M_\odot$, the injected embryo manages to grow into a giant planet for all disc sizes, as all embryos experience outward migration and rapid core growth. Further, the final planet mass increases with increasing disc size. For small discs or for low-mass stars, the region of outward migration moves inwards so fast that planets experience very little outward migration and therefore migrate inwards rapidly, halting their growth into a giant. The reduction in the close-in super-Earth fraction reported, for example, by \citet{mulders2015_SEvMstar} and \citet{zink2023_occ_vs_amplitude} can thus potentially be explained if some stars host large, viscously heated protoplanetary discs.

\begin{figure*}
    \centering
    \includegraphics[width = \linewidth]{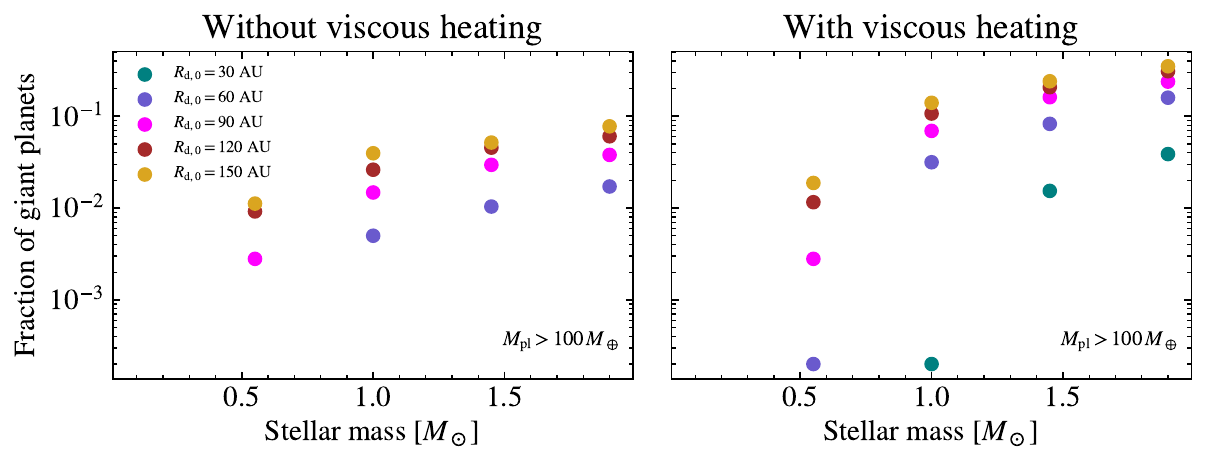}
    \caption{Giant planet fraction at all semi-major axes as a function of stellar mass and disc size. With or without viscous heating, the giant planet fraction increases with increasing stellar mass for all disc sizes where a giant planet can form to begin with. The lower stellar mass limit for giant planet formation is 0.5 $M_\odot$ in both models. In the irradiated case (left), the lower disc size limit for giant planets to form is 60 AU (90 AU for $M_\star = 0.5\, M_\odot$). In contrast, when viscous heating is included, the lower disc size limit is as low as 30 AU for solar-mass stars and above, while it is 60 AU for stars with $M_\star = 0.5\,M_\odot$.}
    \label{fig:scaling_disc_giant_frac}
\end{figure*}

We further show the giant planet fraction as a function of stellar mass for different disc sizes in Fig. \ref{fig:scaling_disc_giant_frac}. We include all giant planets here and not just close-in giants, which means that we divide the number of giants with the total number of injected embryos. This is due to type-II migration being much slower and decreasing with increasing planet mass. As a result, close-in super-Earths become giant planets with final orbits farther away from the star when the size of the disc or the mass of the host star is increased. This can be seen, for example, in figure \ref{fig:alpha_scaling_oneplanet}. The giant planet fraction increases with increasing stellar mass and disc size for both disc models, in agreement with observations, which show that the giant planet occurrence rate peaks for stars with mass $\sim$2 $M_\odot$ \citep{reffert2015_GploccvMstar,ghezzi2018_gpoccmstar}.

Estimated giant planet occurrence rates vary significantly in the literature depending on the mass range and semi-major axis range considered. \citet{fulton2021_giantocc} found that the giant planet occurrence rate for planets within 1-5 AU and masses above 100 $M_\oplus$ ($\sim$0.3 $M_{\rm J}$) is $\sim$15-20\%, while \citet{fernandes2019_GPocc} reported an occurrence rate of 26.6\% for planets between 0.1 and 100 AU and masses of 0.1-20 $M_{\rm J}$. It is important to note that the work of \citet{fernandes2019_GPocc} considers the entire stellar mass range of giant hosts observed by \textit{Kepler}, which includes stars with different masses than the Sun, which can inflate the occurrence rate. \citet{reffert2015_GploccvMstar} investigated the occurrence rates of giant planets as a function of stellar mass and metallicity and found that it increases with increasing stellar mass up to 2 $M_\odot$, with reported occurrence rates ranging from 3-8\% between stellar masses of 1 $M_\odot$ and 2 $M_\odot$. \citet{ghezzi2018_gpoccmstar} found that the giant planet occurrence rate ranges from $\sim$10-20\% in the same stellar mass range. The discrepancy between these two studies is likely caused by the fact that the smallest confirmed minimum mass used by \citet{reffert2015_GploccvMstar} was 2.3 $M_{\rm J}$, while \citet{ghezzi2018_gpoccmstar} considered planets down to masses of 0.5 $M_{\rm J}$. Nevertheless, the observed giant planet occurrence rates for similar mass ranges that we consider seem to vary between 10-20\% for solar mass stars, increasing to at least 20\% for stars with masses of 2 $M_\odot$.

In this work, in the case of irradiation heating only, migration is efficient enough to keep the giant planet fraction low, in the range $\sim$0.5-3\% for solar-mass stars and increases to 2-10\% for stars with masses of 2 $M_\odot$. Discs smaller than 60 AU for all stellar masses fail to form any giants due to their rapid evolution and low masses. When viscous heating is included and inward migration is delayed, the giant planet fraction increases by a factor of $\sim$5 for most stellar masses and protoplanetary disc sizes. These results are in better agreement with observed occurrence rates for giant planets around both solar mass stars and high-mass stars. However, we caution against comparing our planet fractions directly to occurrence rates due to model uncertainties. We discuss the interpretations of our planet fraction more in detail in Sect. \ref{ssec:SE_frac_discussion}.

Further, in the viscous case, giant planets can form even around small discs ($\sim$30 AU) around solar-mass stars and above, which, in our model, is not possible in the irradiated case. Neither model produces any giants for stellar masses of $M_\star \lesssim 0.5
$ $M_\odot$, which is in rough agreement with previous work modelling giant planet formation around varying stellar masses \citep{liu2019_miso,shibatahelled2025_GPform}.

\subsection{Using a fixed injection time}
\label{ssec:same_t0}
In our nominal model, we sampled the injection time uniformly between 10$^3$ yr and the disc lifetime. As a consequence, discs with long lifetimes are negatively affected as planets injected late into the disc (after $\sim$2-3 Myr) struggle to grow to pebble isolation mass. Therefore, we also investigated the effects of different sampling schemes for the injection times.

\begin{figure*}
    \centering
    \includegraphics[width=\linewidth]{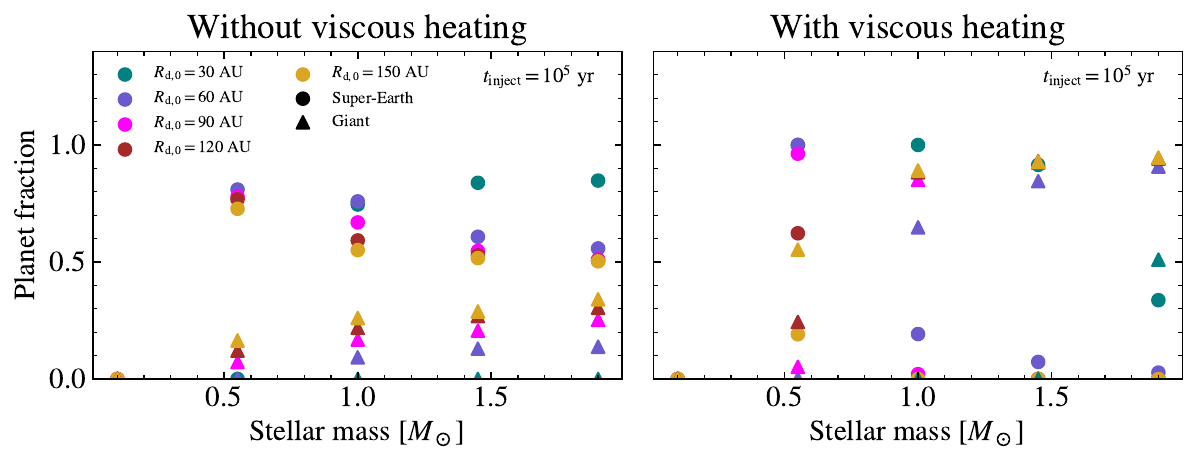}
    \caption{Similar to Fig. \ref{fig:scaling_disc_SE_close_frac} but using a fixed injection time at $t=10^5$ yr. We also show the giant planet fraction (triangles). In both models, super-Earth formation is suppressed for high stellar masses, as planets have time to grow into giant planets instead. As a reminder, the giant planet fraction is defined as the number of giant planets formed compared to the total number of planets injected, while the super-Earth fraction is the number of close-in super-Earths compared to the number of close-in planets, as described in section \ref{ssec:parameter_space}.}
    \label{fig:scaling_disc_SE_close_frac_fixedt0}
\end{figure*}

In Fig. \ref{fig:scaling_disc_SE_close_frac_fixedt0} we show the close-in super-Earth fraction using a fixed injection time at $t=10^5$ yr. We also tested using a fixed injection time at $t=10^3$ yr and $t=10^4$ yr and found no significant differences between the three cases. Further, we tested the effects of sampling the injection time from a log-uniform distribution between each of these times and the disc lifetime and found no significant difference from the case with a fixed injection time. In the irradiated case, we find the opposite trend as in our nominal model. As large discs typically have longer lifetimes and are more massive than small discs, we find that large discs produce fewer super-Earths for a fixed stellar mass as more planets have time to grow into a giant planet instead of a super-Earth. In the case of irradiated heating, the super-Earth fraction decreases beyond 0.5 $M_\odot$ for all disc sizes with the exception of very small discs (30 AU) where the super-Earth fraction continuously increases with increasing stellar mass as we find that no giant planets form for such small discs. For discs larger than 30 AU, we find that the fraction of giant planets increases enough with increasing stellar mass and disc size that the super-Earth fraction is reduced for increasing stellar masses above 0.5 $M_\odot$. For example, for a solar mass star, the giant planet fraction ranges between $\sim$10\% and $\sim$25\% when the injection is fixed for disc sizes between 60 and 150 AU. When sampling injection times throughout the entire disc lifetime, the giant planet fractions instead range between $\sim$0.5\% and $\sim$5\%, which is not high enough to impact the super-Earth fractions.

In the viscous case, we find that the super-Earth fraction is heavily suppressed around large discs while for small discs it peaks at 0.5 $M_\odot$ and decreases significantly for host stars more massive than solar-mass. For discs with sizes of 30 AU, the super-Earth fraction peaks at $M_\star$ = 1.5 $M_\odot$. As all planets are injected early on in the disc lifetime, the outward migration region is located far from the star when the planets start to grow and more planets are affected by some level of outward migration, allowing them to reach higher pebble isolation masses and grow to become giants. The fraction of giant planets approaches unity at solar masses and above for discs larger than 30 AU, which is the cause of the suppressed super-Earth fraction. Further, even the smallest discs are able to form giant planets for high enough stellar masses.

\begin{figure}
    \centering
    \includegraphics[width=\linewidth]{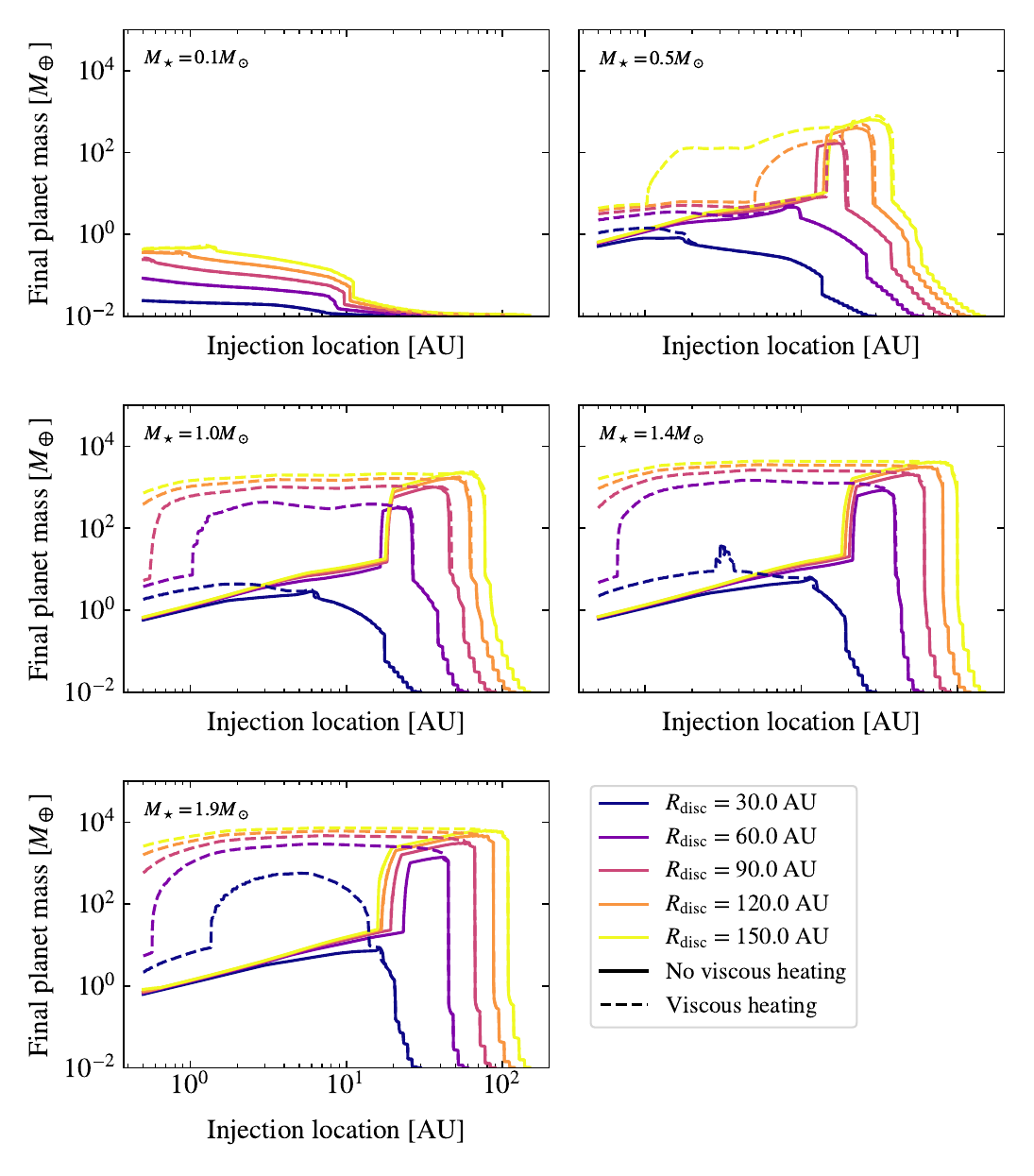}
    \caption{Final planet mass as a function of injection location for different stellar masses and disc sizes, for a fixed injection time of $t=10^5$ yr. For low-mass stars, neither disc model produces any super-Earths or giant planets. In the viscous model, the region where giant planets form moves close to the star for larger discs and stellar masses, resulting in fewer super-Earths forming. In the irradiated case, the giant-forming region remains far out in the disc and only moves slightly closer to the star for higher stellar masses.}
    \label{fig:Mpl_r0}
\end{figure}

For a given system with a fixed injection time, the only parameter determining the final mass of the planet is the injection location. In order for a giant planet to form, the planet needs to be injected sufficiently far out in order for it to be able to reach a large enough pebble isolation mass to accrete a substantial amount of gas, but not too far out as the core will then grow too slowly to reach the pebble isolation mass at all. Planets injected outside of this ‘giant planet region’ will thus not have time to migrate inwards fast enough during the disc lifetime. Planets injected interior of the giant planet region, in contrast, grow to become Earths or super-Earths instead with only minimal gas accretion. To illustrate this distinction, we show the final mass as a function of injection location for the model with fixed injection times in Fig. \ref{fig:Mpl_r0}. In both models, no super-Earths are formed around host stars with a mass of 0.1 $M_\odot$. In the irradiated case for high stellar masses, the final planet mass increases with increasing injection location as planets take longer to migrate to the inner edge and reach higher pebble isolation masses. Planets grow into giants if they are injected far out in the disc at $r\gtrsim$20 AU. This is true for all disc sizes above 30 AU and stellar masses above 0.5 $M_\odot$. For stars with a mass of 0.5 $M_\odot$ and small discs, the final planet mass decreases if the planet is injected beyond 10 AU. 

With viscous heating included, giants start to form for as low stellar masses as 0.5 $M_\odot$ for most disc sizes and injection locations due to outward migration, which allows planet cores to become more massive. The giant planet-forming region covers almost the entire parameter space for disc sizes of 120 AU and above for all stellar masses above 0.1 $M_\odot$, resulting in a severe reduction in super-Earth formation at these disc sizes. For small discs (30 AU), the giant planet-forming region is small for stars with masses of 0.5 $M_\odot$ and 1 $M_\odot$, resulting in high super-Earth fractions while for larger stellar masses, the giant planet-forming region starts to cover the full parameter space, reducing the super-Earth fraction significantly. 

\subsection{Sampling disc sizes and accretion rates}
\label{ssec:sampling}
In the previous section, we scaled the initial accretion rates with stellar mass and explored the $(M_\star,R_{\rm d,0})$-space. This meant that the initial accretion rate was assumed to be fixed for a given stellar mass. However, as seen by, for example, \citet{manara2012_accretion} and \citet{almendros-abad2024_discrelation}, the observed accretion rate exhibits a large scatter around the linear fit. As discussed by \citet{manara2012_accretion}, this scatter might be caused by the observed stars having different ages, as the accretion rate declines over time. Although the scatter remains even when binning the stars by their age, it is therefore possible that the scatter in the accretion rate is either intrinsic or dependent on some other, unknown variable. Further, it is also possible for some of the discs in the parameter space to be gravitationally unstable in some regions of the parameter space. To account for these effects, we turned to a sampling scheme in order to synthesise a more realistic population of discs around stars with different stellar masses and explore the resulting super-Earth fraction. 

\begin{figure*}
    \centering
    \includegraphics[width=\linewidth]{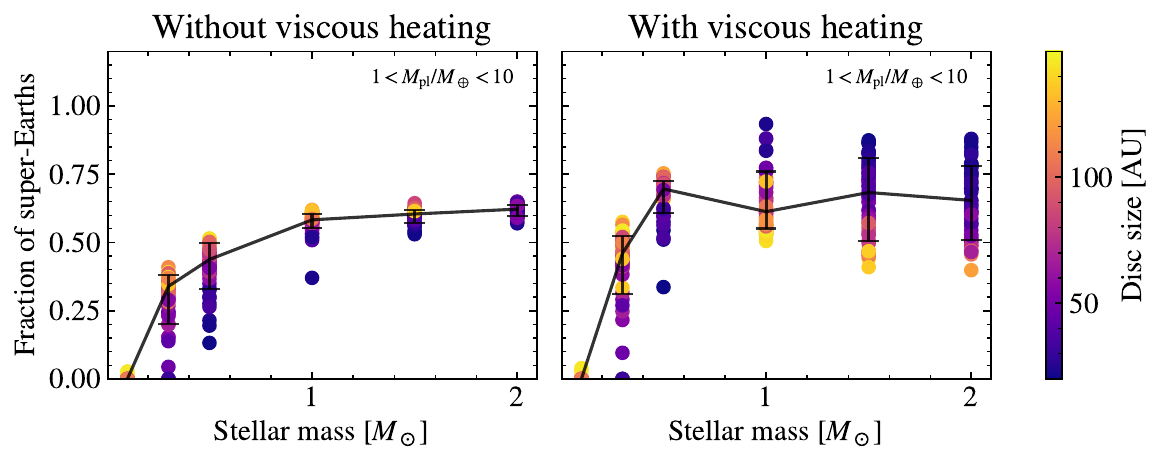}
    \caption{Fraction of close-in super-Earths as a function of stellar mass with sampled initial accretion rates and disc sizes. The black lines show the median values out of all runs, while the error bars show the 16th and 84th percentiles. Similarly to Fig. \ref{fig:scaling_disc_SE_close_frac}, the super-Earth fraction increases with increasing stellar mass in the irradiated case (left). When the disc is viscously heated (right), the super-Earth fraction flattens out for host stars more massive than 0.5 $M_\odot$.}
    \label{fig:sample_disc_SE_close_frac}
\end{figure*}

For a given stellar mass, we sampled the initial accretion rate from a Gaussian centred around the power law in Eq. \eqref{eq:Mdot_Mstar} with a width of $\sigma=0.2$. The disc size was then sampled from a uniform distribution. The minimum disc size was set to 20 AU, while the maximum disc size was set to the smallest value between 150 AU and the largest size at which the entire disc is gravitationally stable. We performed this sampling 50 times. Gravitational stability is determined through the Toomre Q parameter,
\begin{equation}
\label{eq:Q}
    Q = \frac{c_{\rm s} \Omega}{\pi G \Sigma_{\rm g}}.
\end{equation}
Here we calculate $\Sigma_{\rm g}$ and $c_{\rm s}$ using an irradiated temperature profile as $r_{\rm tran} \ll R_{\rm d,0}$. As the disc is gravitationally stable for $Q > 1$, the maximum possible initial disc size for a gravitationally stable disc is given by
\begin{equation}
\label{eq:R_max}
\begin{split}
    R_{\rm d, max,0} & = 173\,{\rm AU}\bigpar{\frac{M_\star}{M_\odot}}^{-1/3}\bigpar{\frac{L_\star}{L_\odot}}^{2/3}\bigpar{\frac{\alphav}{0.01}}^{14/9} \\
    & \times \bigpar{\frac{\dot{M}_{\rm g,0}}{10^{-7}M_\odot/{\rm yr}}}^{-14/9}\bigpar{\frac{\mu_{\rm d}}{2.35}}^{-7/3}\\
    & \approx 173\,{\rm AU}\bigpar{\frac{M_\star}{M_\odot}}^{-1.6}\bigpar{\frac{\alphav}{0.01}}^{14/9}\bigpar{\frac{\mu_{\rm d}}{2.35}}^{-7/3},
\end{split}
\end{equation}
where the last approximation comes from our specific luminosity and gas accretion rate scalings in Eqs. \eqref{eq:L_star} and \eqref{eq:Mdot_Mstar}, respectively. From this, we find that the largest gravitationally stable disc size decreases with increasing stellar mass. Therefore, the range of sampled disc sizes decreases with increasing stellar mass, which means that discs around larger stars are typically smaller than those around low-mass stars when sampled from a uniform distribution. We show the resulting super-Earth fractions in Fig. \ref{fig:sample_disc_SE_close_frac} for both the irradiated and viscous case. As the disc sizes only marginally affect the resulting super-Earth fraction in the irradiated case, we find similar results as when we explored the full parameter space. Due to the sampling of initial accretion rates, however, we find that there is a greater variance in the resulting super-Earth fraction around low-mass stars, with small discs and low initial accretion rates resulting in slightly lower super-Earth fractions.

In the viscous case, we find that the super-Earth fraction peaks at $\sim$0.5 $M_\odot$ and plateaus for higher stellar masses, although the variance is large for all stellar masses. As the discs around high-mass stars are typically smaller, the super-Earth fraction remains relatively high for high stellar masses ($>$$1\,M_\odot$) since the viscously heated region in small discs moves inwards faster than around large discs, resulting in more efficient inward migration and fewer giant planets forming. Notably, for stellar masses of 0.5 $M_\odot$ and below, the fraction of super-Earths increases with increasing disc size for a given stellar mass while the opposite relationship is found for higher stellar masses. As shown in figure \ref{fig:scaling_disc_giant_frac}, giant planets start to form for stellar masses of 0.5 $M_\odot$ and above. As a result, for higher stellar masses when giant planet formation is possible, increasing the disc size results in even more giant planets forming, reducing the fraction of super-Earths forming. This can also be seen when comparing the right panels of figures \ref{fig:scaling_disc_SE_close_frac} and \ref{fig:scaling_disc_giant_frac}. For stellar masses above 0.5 $M_\odot$, the super-Earth fraction decreases with increasing disc size for a given stellar mass, while the giant planet fraction increases for increasing disc size.

\begin{figure*}
    \centering
    \includegraphics[width=\linewidth]{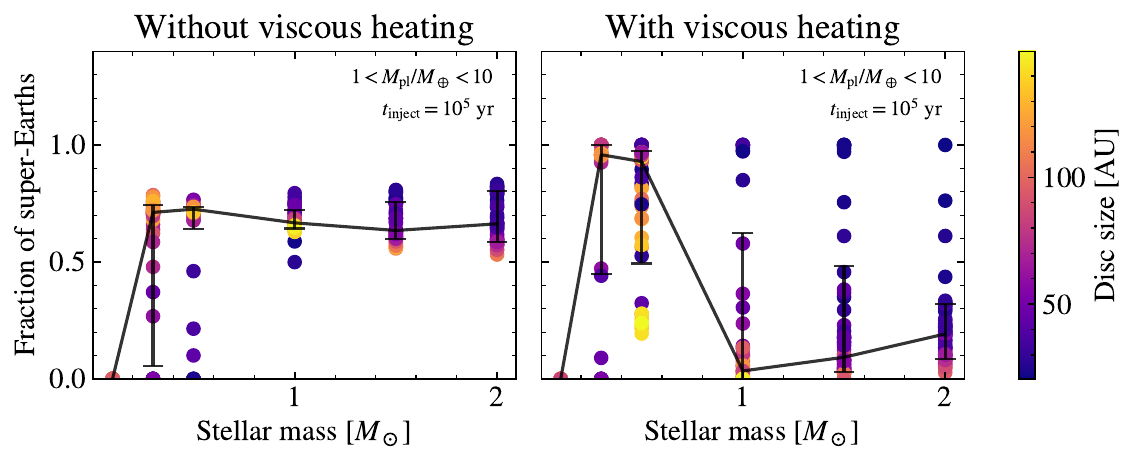}
    \caption{Same as Fig. \ref{fig:sample_disc_SE_close_frac} but with a fixed injection time of 10$^5$ yr. Large discs form more giants when the embryos are injected early, which results in fewer super-Earths being formed. In contrast, if no giants are able to form, such as when the host star mass is below 0.5 $M_\odot$, the super-Earth fraction increases with increasing disc size. We find that the median super-Earth fraction decreases with increasing stellar mass, both with or without viscous heating, although the effect is significantly stronger when viscous heating is active. Further, we find that the median super-Earth fraction increases slightly above 1 $M_\odot$ in the viscous case as the sampled disc sizes around these high-mass stars are slightly smaller.}
    \label{fig:sample_disc_SE_close_frac_fixedt0}
\end{figure*}

In Fig. \ref{fig:sample_disc_SE_close_frac_fixedt0} we show the super-Earth fraction as a function of stellar mass for the sampled discs but using a fixed injection time at $t=10^5$ yr. When the disc is heated purely through irradiation, migration is efficient enough that only protoplanets injected far out in the disc will grow into a giant planet, even when the protoplanets are injected early on in the disc lifetime. Giant planet formation is nevertheless more efficient than in our nominal case due to the early injection time. Therefore, a minor reduction in the median super-Earth fraction with increasing stellar mass is found.

When viscous heating is included, we find similar results as in Fig. \ref{fig:scaling_disc_SE_close_frac_fixedt0} where super-Earth formation is heavily suppressed for most sampled discs, since giant planet formation starts to dominate at solar mass and above. For large discs around low-mass stars, we also find that giant planet formation becomes efficient enough to dominate over super-Earth formation. For most sampled discs, no matter where the planet is injected, it has enough time to grow into a giant planet as outward migration is efficient in the early lifetimes of these discs, resulting in fewer super-Earths formed.

As mentioned, the sample range of disc sizes around high-mass stars is smaller than that of low-mass stars due to gravitational instability, which leads to smaller disc sizes. As a result, the median super-Earth fraction increases slightly around stars with masses of $\sim$2 $M_\odot$. However, this increase is marginal as even small discs predominately form giant planets when viscous heating is included and a fixed, early, injection time is used.
\section{Discussion}
\label{sec:discussion}

\subsection{Comparing with previous work}
\label{ssec:prev_work}
Considering planetesimal accretion as the main planet formation pathway, \citet{burn2021_lowmassstars} performed planet population synthesis for stars with masses below solar mass. They found that giant planet formation occurs for stellar masses of at least 0.5 $M_\odot$ and that the occurrence rates of giant planets increase with increasing stellar masses, in line with the results we have found in this work. Further, they found that the super-Earth occurrence rate increased with increasing stellar mass up to masses of $\sim$0.5 $M_\odot$, after which the trend plateaus, despite including outward migration in their work. These discrepancies compared to our work could be due, for example, to their relatively low disc masses, which result in less efficient giant planet formation and consequently less inhibition of super-Earth formation. Moreover, giant planet formation has been found to be less efficient for planetesimal accretion models compared to pebble accretion, especially in the outer regions of the protoplanetary disc, \citep[e.g.][]{lorek_johansen2022_planetesimalacc}. If giant planet formation is inefficient, super-Earth formation never gets suppressed and continues to increase with increasing stellar mass. 
\\\\
Recently, \citet{pan2025_SEvMstar} performed planet population synthesis using pebble accretion and similar disc models to this work, including viscous heating and outward migration. They found that the occurrence rates of super-Earths increased for increasing stellar masses up to $\sim$0.5 $M_\odot$, after which it decreases, similar to the models in this work where viscous heating and outward migration is included. Further, they found that giant planets could still form for stellar masses below 0.5 $M_\odot$, in contrast to this work and the work of \citet{burn2021_lowmassstars}, due to the effects of efficient pebble accretion and planet-planet collisions, which aids in the formation of massive planetary cores.
\subsection{Interpreting planet fractions}
\label{ssec:SE_frac_discussion}
Throughout this work, we determined the fraction of planets by calculating the number of planets in our defined mass ranges compared to the total number of planets injected. While it is tempting to try to compare these values with observed occurrence rates of super-Earths and giant planets, we caution against this due to large uncertainties in the choice of model and simulation setup. 
\\\\
One major reason for this caution is that the final planet fractions are strongly affected by how the injection time and location are sampled. For instance, in our nominal model without any viscous heating, the giant planet fraction varies between $\sim$0.5\%-5\% and $\sim$2\%-10\% for stars with masses of 1 $M_\odot$ and 2 $M_\odot$, respectively. This can be contrasted to when we fix the injection time to $10^3$ yr, where the giant planet fraction varies from $\sim$10-25\% up to $\sim$20-35\%.
\\\\
Further, our super-Earth fractions in particular are sensitive to our model choices, especially the lack of N-body interactions, which we neglected in favour of computational speed. Including multiple planets in our model could cause the planetary system to become unstable and super-Earths to be ejected when, for example, a giant planet migrates inwards to the inner regions of the disc \citep[e.g.][]{mustill2015_HJclearingSE,bitschizidoro2023_giantbullies}. For further discussion on this, see Sect. \ref{ssec:N_body}. 
\\\\
As a result, there is no direct translation possible between our resulting planet fraction and observed occurrence rates of both giant planets and super-Earths. Instead, our resulting planet fractions should be interpreted primarily as the probability that a certain type of planet forms in isolation around a given star, given a set of initial conditions and model assumptions, rather than the fraction of stars that host that type of planet.
\\\\
Despite this, we argue that the trends for the dependence of our planet fractions with respect to stellar mass give useful insight into the underlying causes for the observed occurrence rate. For example, a lower super-Earth fraction does indicate that super-Earths are less likely to form, which in turn would result in a reduced occurrence rate, but the absolute values cannot necessarily be directly converted into each other.

\subsection{Viscous heating of discs}
\label{ssec:visc_heating}
Whether or not discs are viscously heated is uncertain. When non-ideal magnetohydrodynamics (MHD) are taken into account, simulations have found that viscous heating mostly takes place at several scale heights above the midplane as the magnetorotational instability (MRI) becomes suppressed in the midplane \citep{mori2019_viscousheating}. In these cases, the midplane temperature is dominated by stellar irradiation. As shown in this work, the temperature structure of the disc can have a significant impact on the migration of planets in the disc, and understanding to what extent discs are viscously heated is of great significance in order to understand the properties and origins of the resulting planet population. 
\\\\
When we assume that viscous heating takes place in the midplane, the two main parameters that determine the temperature are $\alphav$ and $\dot{M}_{\rm g}$, which are both related to each other. The value of $\alphav$ for discs is currently estimated to be between $10^{-4}$ and $10^{-2}$ \citep{pinte2016_alpha,najitabergin2018_alpha, trapman2020_alpha,villenave2022_alpha,rosotti2023_alphaobs}. For lower values of $\alphav$, the initial accretion rate also needs to be lower in order to achieve realistic disc sizes for a fixed disc mass. For example, for $\alphav = 10^{-3}$ and $\dot{M}_{\rm g,0} = 10^{-7}\,M_\odot/{\rm yr}$, the resulting initial disc size is $\sim$6 AU for a fixed disc mass of 0.1 $M_\odot$, which we consider unrealistic as dust disc sizes of young discs have been observed to be $\gtrsim$10 AU \citep{ohashi2023_edisk} and are expected to be smaller than the gas disc due to pebble drift \citep[e.g.][]{ansdell2018_dustsizes}. If instead $\dot{M}_{\rm g,0} = 10^{-8}\,M_\odot/{\rm yr}$, the disc size remains at 71 AU. In order to test the effects of $\alphav$, we ran the same simulations as in Sect. \ref{ssec:parameter_space} but with $\alphav = 10^{-3}$ and $\dot{M}_{\rm g,0} = 10^{-8}\,M_\odot/{\rm yr}$. We show the resulting super-Earth fractions in Fig. \ref{fig:scaling_disc_lowalpha} in Appendix \ref{app:additional_figs} and find that the resulting super-Earth fractions are qualitatively similar to our nominal model. In the low viscosity case, the transition radius, $r_{\rm tran}$, between the viscously heated region and the irradiated region at $t=0$ is closer to the star at 1.8 AU compared to 4.9 AU in our nominal model. This is a consequence of the lower accretion rates, which results in weaker viscous heating. When the transition radius is closer to the star, inward migration is more efficient and the reduction of close-in super-Earths is less significant around high-mass stars. However, the initial surface density at $t=0$ is also higher in the viscous regime, meaning outward migration is more efficient. This slightly reduces the effects of the efficient inward migration and we find that the fraction of close-in super-Earths still decreases with increasing stellar masses for large, viscously heated discs. Further, migration (both inwards and outwards) in both the irradiated and viscous case remains fast for a longer time for low $\alphav$, since the surface density of the low viscosity disc remains high over a longer timescale. As a result, inward migration is more efficient in irradiated discs and the fraction of close-in super-Earths peaks at a stellar mass of $\sim$0.5 $M_\odot$ and decreases only slightly for higher stellar masses as migration is efficient enough that only a few planets grow beyond the super-Earth range. This is found for all disc sizes as disc size does not have a significant effect in the irradiated case.

\subsection{Varying fragmentation velocities}
\label{ssec:v_f}
Throughout this work, we used a constant fragmentation velocity of 2 m/s throughout the entire disc. This resulted in St $\sim$ 0.01 in the outer disc, which is consistent with observations \citep{jiang2024_dustsizes}. However, fragmentation velocities of dust in protoplanetary discs are highly uncertain, with experiments finding fragmentation velocities to vary between 1 m/s and 10 m/s depending on the composition of dust \citep[e.g.][]{gundlach_blum2015_vfrag}. Previous work has therefore often chosen to vary the fragmentation velocity such that dry pebbles inside the water iceline have fragmentation velocities of 1 m/s while icy pebbles outside the water iceline have fragmentation velocities of 10 m/s \citep[e.g.][]{drazkowska2017_snowline,venturini2020_water}. In previous work, \citet{nielsen2025} showed that using a varying fragmentation velocity had little to no effect on the super-Earth population around a solar-mass star. However, considering that the location of the iceline is dependent on the temperature of the disc and therefore on the stellar mass, it is possible that varying the fragmentation velocity might play a bigger role around stars with different stellar masses. We therefore tested the effects of different fragmentation velocities on the close-in super-Earth population. We ran the same set of simulations as in Sect. \ref{ssec:parameter_space} but with fragmentation velocities of 1 m/s and 10 m/s for dry and icy particles, respectively. We show the results in Fig. \ref{fig:scaling_disc_varied_vf} in Appendix \ref{app:additional_figs}. In the irradiated case, the results are relatively unchanged while in the viscous case, the super-Earth fraction is even more reduced for higher stellar masses for all disc sizes. This is a result of the water iceline moving outwards for higher stellar masses, which in turn results in the inefficient, dry region, with small pebbles, extending to larger parts of the disc, resulting in fewer super-Earths being formed. Further, the giant planet formation beyond the water iceline is enhanced due to the larger pebbles, which in turn reduces the super-Earth fraction farther. 

\subsection{Including N-body effects}
\label{ssec:N_body}
In this work we only included one protoplanet per disc, and the injected planets do not interact with each other. Including N-body effects could have a significant impact on the final planet population, as planet-planet collisions have shown to be able, for example, to produce giant planets around low-mass stars \citep{pan2024_collisions}. As multiple super-Earths migrate inwards through the disc, they can get trapped in resonant chains, which become unstable over time and break, causing planet-planet collisions \citep[e.g.][]{izidoro2021_breakingchains,esteves2022_breakingchains,kajtazi2023_resonance}. Given that the Hill radii of planets of a given mass and semi-major axis are higher around low-mass stars, it is therefore natural to expect that planet-planet collisions can be more prevalent around low-mass stars. Further, as inward migration is efficient around low-mass stars, especially in irradiated discs, the enhanced close-in super-Earth fraction could be explained by efficient inward migration and subsequent collisions \citep[e.g.][]{pan2025_SEvMstar}. 
\\\\
Due to their shorter lifetimes on the main-sequence, most high-mass stars that are observed today are young. From galactic chemical evolution, young stars are typically more metal-rich \citep{feuillet2018_agefeh,delgadomena2019_agefeh}. As a result, observed high-mass stars are on average more metal-rich \citep[see e.g. fig. 2 in ][]{nielsen2023}, and are more likely to host giant planets. The presence of inner super-Earths has been found to be positively correlated with the presence of an outer giant for stars with super-solar mass and super-solar metallicity \citep{bryanlee2024_innerSE_outerGP,bryanlee2025_innerSPouterGP,bonomo2025_innerSEouterGP}. This indicates that if a single giant planet forms and remains on a wide orbit, any potential inner planets typically remain undisturbed by the giant. However, if instead multiple giants form in the system, these can cause instabilities, which in turn can eject inner super-Earths from the system, provided that the giants form or migrate close enough to the super-Earths \citep{mustill2015_HJclearingSE,bitschizidoro2023_giantbullies,kong2024_giantbully}. Therefore, in order to understand the effects of stellar mass on the inner super-Earth population, it will be necessary to consider the formation of multiple planets and their gravitational interactions in a given system.

\section{Conclusions}
\label{sec:conclusion}
In this work, we simulated the formation of planets around different stellar masses using two different disc models with either purely irradiated heating or both viscous and irradiated heating. We find the following:
\begin{enumerate}
    \item When the disc is heated purely through irradiation, type-I migration is exclusively inwards and very efficient, resulting in rapid inward planet migration and an increase in the close-in super-Earth fraction for higher stellar masses These results are roughly independent of disc size. \\
    \item When viscous heating is included, the direction of type-I migration is flipped in the inner disc for a range of relevant planet masses, which delays inward migration slightly. This delay results in a significant fraction of planets growing into giants instead. The transition radius between inwards and outward migration is dependent on stellar mass and moves closer to the star over time. Around small discs or low-mass stars, this region of outward migration moves close to the star quickly enough that planets experience rapid inward migration, which halts their growth into giants. In contrast, larger discs evolve over a longer timescale, which results in inward migration being delayed for longer, allowing more giants planets to form. \\
    \item As discs with larger sizes or those around high-mass stars have longer lifetimes, giant planet formation is negatively affected when sampling injection times over the entire disc lifetime. This is a result of planet growth being inefficient in the later stages of the disc. When fixing the injection time to $10^{5}$ yr, we find that the super-Earth fraction is reduced around high-mass stars in both the irradiated and viscous case. In the irradiated case, more planets around high-mass stars grow into giants compared to when sampling injection times, resulting in a slightly reduced super-Earth fraction around high-mass stars. In the viscous case, a majority of planets around solar-mass stars and above grow into giants when injected this early due to outward migration, resulting in a highly suppressed super-Earth fraction around high-mass stars and in large discs. We also tested sampling the injection time from a log-uniform distribution and found little difference compared to using a fixed injection time. \\
    \item When sampling disc parameters from a distribution, discs are typically smaller around high-mass stars. As a result, slightly fewer giants form around high-mass stars when sampling disc parameters compared to the case when disc sizes increase with increasing stellar mass. However, we still find that the super-Earth fraction is reduced around high-mass stars when viscous heating is included. \\
    \item We tested our results for lower values of $\alphav$ and different fragmentation velocities and find that the fraction of super-Earths around high-mass stars is still reduced when viscous heating is active, similar to our nominal model.
\end{enumerate}
This work shows how stellar masses and the structure of protoplanetary discs can shape the resulting planetary systems. In particular, we find that the reduction of close-in super-Earths around high-mass stars can be explained by an increase in giant planet formation efficiency around more massive stars. Future statistics of planetary systems around a wide range of stellar masses as well as better constraints on the structure of protoplanetary discs could help confirm these predictions.

\begin{acknowledgements}
\textit{We thank the anonymous referee for their comments that helped improve this manuscript. A.J. acknowledges funding from, the Carlsberg Foundation (Semper Ardens: Advance grant FIRSTATMO), and the Göran Gustafsson Foundation.}
\end{acknowledgements}
\bibliographystyle{aa}
\bibliography{ref}

\begin{thebibliography}{89}
\expandafter\ifx\csname natexlab\endcsname\relax\def\natexlab#1{#1}\fi

\bibitem[{{Alibert} {et~al.}(2018){Alibert}, {Venturini}, {Helled}, {Ataiee}, {Burn}, {Senecal}, {Benz}, {Mayer}, {Mordasini}, {Quanz}, \& {Sch{\"o}nb{\"a}chler}}]{alibert2018_hybridmodel}
{Alibert}, Y., {Venturini}, J., {Helled}, R., {et~al.} 2018, Nature Astronomy, 2, 873

\bibitem[{{Almendros-Abad} {et~al.}(2024){Almendros-Abad}, {Manara}, {Testi}, {Natta}, {Claes}, {Mu{\v{z}}i{\'c}}, {Sanchis}, {Alcal{\'a}}, {Bayo}, \& {Scholz}}]{almendros-abad2024_discrelation}
{Almendros-Abad}, V., {Manara}, C.~F., {Testi}, L., {et~al.} 2024, \aap, 685, A118

\bibitem[{{Ansdell} {et~al.}(2018){Ansdell}, {Williams}, {Trapman}, {van Terwisga}, {Facchini}, {Manara}, {van der Marel}, {Miotello}, {Tazzari}, {Hogerheijde}, {Guidi}, {Testi}, \& {van Dishoeck}}]{ansdell2018_dustsizes}
{Ansdell}, M., {Williams}, J.~P., {Trapman}, L., {et~al.} 2018, \apj, 859, 21

\bibitem[{{Bae} {et~al.}(2013){Bae}, {Hartmann}, {Zhu}, \& {Gammie}}]{bae2013_xray}
{Bae}, J., {Hartmann}, L., {Zhu}, Z., \& {Gammie}, C. 2013, \apj, 774, 57

\bibitem[{{Baraffe} {et~al.}(2002){Baraffe}, {Chabrier}, {Allard}, \& {Hauschildt}}]{baraffe2002_lum}
{Baraffe}, I., {Chabrier}, G., {Allard}, F., \& {Hauschildt}, P.~H. 2002, \aap, 382, 563

\bibitem[{{Baraffe} {et~al.}(2015){Baraffe}, {Homeier}, {Allard}, \& {Chabrier}}]{baraffe2015_lum}
{Baraffe}, I., {Homeier}, D., {Allard}, F., \& {Chabrier}, G. 2015, \aap, 577, A42

\bibitem[{{Bell} \& {Lin}(1994)}]{bell_lin1994_opacities}
{Bell}, K.~R. \& {Lin}, D.~N.~C. 1994, \apj, 427, 987

\bibitem[{{Ben{\'\i}tez-Llambay} \& {Pessah}(2018)}]{benitez2018_dusttorques}
{Ben{\'\i}tez-Llambay}, P. \& {Pessah}, M.~E. 2018, \apjl, 855, L28

\bibitem[{{Bitsch} \& {Izidoro}(2023)}]{bitschizidoro2023_giantbullies}
{Bitsch}, B. \& {Izidoro}, A. 2023, \aap, 674, A178

\bibitem[{{Bitsch} {et~al.}(2015{\natexlab{a}}){Bitsch}, {Johansen}, {Lambrechts}, \& {Morbidelli}}]{bitsch2015_giants}
{Bitsch}, B., {Johansen}, A., {Lambrechts}, M., \& {Morbidelli}, A. 2015{\natexlab{a}}, \aap, 575, A28

\bibitem[{{Bitsch} {et~al.}(2015{\natexlab{b}}){Bitsch}, {Lambrechts}, \& {Johansen}}]{bitsch2015b_gasaccretion}
{Bitsch}, B., {Lambrechts}, M., \& {Johansen}, A. 2015{\natexlab{b}}, \aap, 582, A112

\bibitem[{{Bitsch} {et~al.}(2018){Bitsch}, {Morbidelli}, {Johansen}, {Lega}, {Lambrechts}, \& {Crida}}]{bitsch2018_Miso}
{Bitsch}, B., {Morbidelli}, A., {Johansen}, A., {et~al.} 2018, \aap, 612, A30

\bibitem[{{Bonomo} {et~al.}(2025){Bonomo}, {Naponiello}, {Pezzetta}, {Sozzetti}, {Gandolfi}, {Wittenmyer}, \& {Pinamonti}}]{bonomo2025_innerSEouterGP}
{Bonomo}, A.~S., {Naponiello}, L., {Pezzetta}, E., {et~al.} 2025, \aap, 700, A126

\bibitem[{{Borucki} {et~al.}(2010){Borucki}, {Koch}, {Basri}, {Batalha}, {Brown}, {Caldwell}, {Caldwell}, {Christensen-Dalsgaard}, {Cochran}, {DeVore}, {Dunham}, {Dupree}, {Gautier}, {Geary}, {Gilliland}, {Gould}, {Howell}, {Jenkins}, {Kondo}, {Latham}, {Marcy}, {Meibom}, {Kjeldsen}, {Lissauer}, {Monet}, {Morrison}, {Sasselov}, {Tarter}, {Boss}, {Brownlee}, {Owen}, {Buzasi}, {Charbonneau}, {Doyle}, {Fortney}, {Ford}, {Holman}, {Seager}, {Steffen}, {Welsh}, {Rowe}, {Anderson}, {Buchhave}, {Ciardi}, {Walkowicz}, {Sherry}, {Horch}, {Isaacson}, {Everett}, {Fischer}, {Torres}, {Johnson}, {Endl}, {MacQueen}, {Bryson}, {Dotson}, {Haas}, {Kolodziejczak}, {Van Cleve}, {Chandrasekaran}, {Twicken}, {Quintana}, {Clarke}, {Allen}, {Li}, {Wu}, {Tenenbaum}, {Verner}, {Bruhweiler}, {Barnes}, \& {Prsa}}]{borucki2010_kepler}
{Borucki}, W.~J., {Koch}, D., {Basri}, G., {et~al.} 2010, Science, 327, 977

\bibitem[{{Bryan} \& {Lee}(2024)}]{bryanlee2024_innerSE_outerGP}
{Bryan}, M.~L. \& {Lee}, E.~J. 2024, \apjl, 968, L25

\bibitem[{{Bryan} \& {Lee}(2025)}]{bryanlee2025_innerSPouterGP}
{Bryan}, M.~L. \& {Lee}, E.~J. 2025, \apjl, 982, L7

\bibitem[{{Burn} {et~al.}(2021){Burn}, {Schlecker}, {Mordasini}, {Emsenhuber}, {Alibert}, {Henning}, {Klahr}, \& {Benz}}]{burn2021_lowmassstars}
{Burn}, R., {Schlecker}, M., {Mordasini}, C., {et~al.} 2021, \aap, 656, A72

\bibitem[{{Chachan} \& {Lee}(2023)}]{chachan2023_SE_Mdwarf}
{Chachan}, Y. \& {Lee}, E.~J. 2023, \apjl, 952, L20

\bibitem[{{D'Angelo} \& {Lubow}(2010)}]{dangelo_lubow2010}
{D'Angelo}, G. \& {Lubow}, S.~H. 2010, \apj, 724, 730

\bibitem[{{Danti} {et~al.}(2025){Danti}, {Lambrechts}, \& {Lorek}}]{danti2025_SEformGP}
{Danti}, C., {Lambrechts}, M., \& {Lorek}, S. 2025, \aap, 700, A132

\bibitem[{{Delgado Mena} {et~al.}(2019){Delgado Mena}, {Moya}, {Adibekyan}, {Tsantaki}, {Gonz{\'a}lez Hern{\'a}ndez}, {Israelian}, {Davies}, {Chaplin}, {Sousa}, {Ferreira}, \& {Santos}}]{delgadomena2019_agefeh}
{Delgado Mena}, E., {Moya}, A., {Adibekyan}, V., {et~al.} 2019, \aap, 624, A78

\bibitem[{{Dr{\k{a}}{\.z}kowska} \& {Alibert}(2017)}]{drazkowska2017_snowline}
{Dr{\k{a}}{\.z}kowska}, J. \& {Alibert}, Y. 2017, \aap, 608, A92

\bibitem[{{Dr{\k{a}}{\.z}kowska} {et~al.}(2023){Dr{\k{a}}{\.z}kowska}, {Bitsch}, {Lambrechts}, {Mulders}, {Harsono}, {Vazan}, {Liu}, {Ormel}, {Kretke}, \& {Morbidelli}}]{drazkowska2023_pfreview}
{Dr{\k{a}}{\.z}kowska}, J., {Bitsch}, B., {Lambrechts}, M., {et~al.} 2023, in Astronomical Society of the Pacific Conference Series, Vol. 534, Protostars and Planets VII, ed. S.~{Inutsuka}, Y.~{Aikawa}, T.~{Muto}, K.~{Tomida}, \& M.~{Tamura}, 717

\bibitem[{{Dr{\k{a}}{\.z}kowska} {et~al.}(2021){Dr{\k{a}}{\.z}kowska}, {Stammler}, \& {Birnstiel}}]{drazkowska2021_pebbleflux}
{Dr{\k{a}}{\.z}kowska}, J., {Stammler}, S.~M., \& {Birnstiel}, T. 2021, \aap, 647, A15

\bibitem[{{Ercolano} {et~al.}(2023){Ercolano}, {Picogna}, \& {Monsch}}]{ercolano2023_discPE}
{Ercolano}, B., {Picogna}, G., \& {Monsch}, K. 2023, \mnras, 526, L105

\bibitem[{{Ercolano} {et~al.}(2021){Ercolano}, {Picogna}, {Monsch}, {Drake}, \& {Preibisch}}]{ercolano2021_PELx}
{Ercolano}, B., {Picogna}, G., {Monsch}, K., {Drake}, J.~J., \& {Preibisch}, T. 2021, \mnras, 508, 1675

\bibitem[{{Esteves} {et~al.}(2022){Esteves}, {Izidoro}, {Bitsch}, {Jacobson}, {Raymond}, {Deienno}, \& {Winter}}]{esteves2022_breakingchains}
{Esteves}, L., {Izidoro}, A., {Bitsch}, B., {et~al.} 2022, \mnras, 509, 2856

\bibitem[{{Fernandes} {et~al.}(2019){Fernandes}, {Mulders}, {Pascucci}, {Mordasini}, \& {Emsenhuber}}]{fernandes2019_GPocc}
{Fernandes}, R.~B., {Mulders}, G.~D., {Pascucci}, I., {Mordasini}, C., \& {Emsenhuber}, A. 2019, \apj, 874, 81

\bibitem[{{Feuillet} {et~al.}(2018){Feuillet}, {Bovy}, {Holtzman}, {Weinberg}, {Garc{\'\i}a-Hern{\'a}ndez}, {Hearty}, {Majewski}, {Roman-Lopes}, {Rybizki}, \& {Zamora}}]{feuillet2018_agefeh}
{Feuillet}, D.~K., {Bovy}, J., {Holtzman}, J., {et~al.} 2018, \mnras, 477, 2326

\bibitem[{{Fulton} {et~al.}(2021){Fulton}, {Rosenthal}, {Hirsch}, {Isaacson}, {Howard}, {Dedrick}, {Sherstyuk}, {Blunt}, {Petigura}, {Knutson}, {Behmard}, {Chontos}, {Crepp}, {Crossfield}, {Dalba}, {Fischer}, {Henry}, {Kane}, {Kosiarek}, {Marcy}, {Rubenzahl}, {Weiss}, \& {Wright}}]{fulton2021_giantocc}
{Fulton}, B.~J., {Rosenthal}, L.~J., {Hirsch}, L.~A., {et~al.} 2021, \apjs, 255, 14

\bibitem[{{Ghezzi} {et~al.}(2018){Ghezzi}, {Montet}, \& {Johnson}}]{ghezzi2018_gpoccmstar}
{Ghezzi}, L., {Montet}, B.~T., \& {Johnson}, J.~A. 2018, \apj, 860, 109

\bibitem[{{Guilera} {et~al.}(2025){Guilera}, {Benitez-Llambay}, {Miller Bertolami}, \& {Pessah}}]{guilera2025_dusttorques}
{Guilera}, O.~M., {Benitez-Llambay}, P., {Miller Bertolami}, M.~M., \& {Pessah}, M.~E. 2025, \apj, 986, 199

\bibitem[{{Gundlach} \& {Blum}(2015)}]{gundlach_blum2015_vfrag}
{Gundlach}, B. \& {Blum}, J. 2015, \apj, 798, 34

\bibitem[{{Gurrutxaga} {et~al.}(2024){Gurrutxaga}, {Johansen}, {Lambrechts}, \& {Appelgren}}]{gurrutxaga2023_outergiant}
{Gurrutxaga}, N., {Johansen}, A., {Lambrechts}, M., \& {Appelgren}, J. 2024, \aap, 682, A43

\bibitem[{{Hartmann} {et~al.}(1998){Hartmann}, {Calvet}, {Gullbring}, \& {D'Alessio}}]{hartmann1998}
{Hartmann}, L., {Calvet}, N., {Gullbring}, E., \& {D'Alessio}, P. 1998, \apj, 495, 385

\bibitem[{{Hartmann} {et~al.}(2016){Hartmann}, {Herczeg}, \& {Calvet}}]{hartmann2016}
{Hartmann}, L., {Herczeg}, G., \& {Calvet}, N. 2016, \araa, 54, 135

\bibitem[{{Haugb{\o}lle} {et~al.}(2019){Haugb{\o}lle}, {Weber}, {Wielandt}, {Ben{\'\i}tez-Llambay}, {Bizzarro}, {Gressel}, \& {Pessah}}]{haugbolle2019_filtering}
{Haugb{\o}lle}, T., {Weber}, P., {Wielandt}, D.~P., {et~al.} 2019, \aj, 158, 55

\bibitem[{{Huang} {et~al.}(2025){Huang}, {Yu}, {Lee}, {Dong}, \& {Bai}}]{huang2025_filtering}
{Huang}, P., {Yu}, F., {Lee}, E.~J., {Dong}, R., \& {Bai}, X.-N. 2025, \apj, 988, 94

\bibitem[{{Ida} {et~al.}(2016){Ida}, {Guillot}, \& {Morbidelli}}]{ida2016}
{Ida}, S., {Guillot}, T., \& {Morbidelli}, A. 2016, \aap, 591, A72

\bibitem[{{Ida} {et~al.}(2018){Ida}, {Tanaka}, {Johansen}, {Kanagawa}, \& {Tanigawa}}]{ida2018}
{Ida}, S., {Tanaka}, H., {Johansen}, A., {Kanagawa}, K.~D., \& {Tanigawa}, T. 2018, \apj, 864, 77

\bibitem[{{Izidoro} {et~al.}(2021){Izidoro}, {Bitsch}, {Raymond}, {Johansen}, {Morbidelli}, {Lambrechts}, \& {Jacobson}}]{izidoro2021_breakingchains}
{Izidoro}, A., {Bitsch}, B., {Raymond}, S.~N., {et~al.} 2021, \aap, 650, A152

\bibitem[{{Jiang} {et~al.}(2024){Jiang}, {Mac{\'\i}as}, {Guerra-Alvarado}, \& {Carrasco-Gonz{\'a}lez}}]{jiang2024_dustsizes}
{Jiang}, H., {Mac{\'\i}as}, E., {Guerra-Alvarado}, O.~M., \& {Carrasco-Gonz{\'a}lez}, C. 2024, \aap, 682, A32

\bibitem[{{Johansen} {et~al.}(2019){Johansen}, {Ida}, \& {Brasser}}]{johansen2019}
{Johansen}, A., {Ida}, S., \& {Brasser}, R. 2019, \aap, 622, A202

\bibitem[{{Johansen} {et~al.}(2021){Johansen}, {Ronnet}, {Bizzarro}, {Schiller}, {Lambrechts}, {Nordlund}, \& {Lammer}}]{johansen2021_pebSS}
{Johansen}, A., {Ronnet}, T., {Bizzarro}, M., {et~al.} 2021, Science Advances, 7, eabc0444

\bibitem[{{Johnson} {et~al.}(2010){Johnson}, {Aller}, {Howard}, \& {Crepp}}]{johnson2010_giantocc}
{Johnson}, J.~A., {Aller}, K.~M., {Howard}, A.~W., \& {Crepp}, J.~R. 2010, \pasp, 122, 905

\bibitem[{{Kajtazi} {et~al.}(2023){Kajtazi}, {Petit}, \& {Johansen}}]{kajtazi2023_resonance}
{Kajtazi}, K., {Petit}, A.~C., \& {Johansen}, A. 2023, \aap, 669, A44

\bibitem[{{Kanagawa} {et~al.}(2018){Kanagawa}, {Tanaka}, \& {Szuszkiewicz}}]{kanagawa2018}
{Kanagawa}, K.~D., {Tanaka}, H., \& {Szuszkiewicz}, E. 2018, \apj, 861, 140

\bibitem[{{Kley} \& {Nelson}(2012)}]{kleynelson2012_migration}
{Kley}, W. \& {Nelson}, R.~P. 2012, \araa, 50, 211

\bibitem[{{Kong} {et~al.}(2024){Kong}, {Johansen}, {Lambrechts}, {Jiang}, \& {Zhu}}]{kong2024_giantbully}
{Kong}, Z., {Johansen}, A., {Lambrechts}, M., {Jiang}, J.~H., \& {Zhu}, Z.-H. 2024, \aap, 687, A121

\bibitem[{{Lef{\`e}vre-Forj{\'a}n} \& {Mulders}(2025)}]{lefevremulders2025_innerSEoutersaturn}
{Lef{\`e}vre-Forj{\'a}n}, E. \& {Mulders}, G.~D. 2025, \apj, 988, 101

\bibitem[{{Lesur} {et~al.}(2023){Lesur}, {Flock}, {Ercolano}, {Lin}, {Yang}, {Barranco}, {Benitez-Llambay}, {Goodman}, {Johansen}, {Klahr}, {Laibe}, {Lyra}, {Marcus}, {Nelson}, {Squire}, {Simon}, {Turner}, {Umurhan}, \& {Youdin}}]{lesur2023_pp7discreview}
{Lesur}, G., {Flock}, M., {Ercolano}, B., {et~al.} 2023, in Astronomical Society of the Pacific Conference Series, Vol. 534, Protostars and Planets VII, ed. S.~{Inutsuka}, Y.~{Aikawa}, T.~{Muto}, K.~{Tomida}, \& M.~{Tamura}, 465

\bibitem[{{Liu} {et~al.}(2019){Liu}, {Lambrechts}, {Johansen}, \& {Liu}}]{liu2019_miso}
{Liu}, B., {Lambrechts}, M., {Johansen}, A., \& {Liu}, F. 2019, \aap, 632, A7

\bibitem[{{Lorek} \& {Johansen}(2022)}]{lorek_johansen2022_planetesimalacc}
{Lorek}, S. \& {Johansen}, A. 2022, \aap, 666, A108

\bibitem[{{Lyra} {et~al.}(2023){Lyra}, {Johansen}, {Ca{\~n}as}, \& {Yang}}]{lyra2023_growth}
{Lyra}, W., {Johansen}, A., {Ca{\~n}as}, M.~H., \& {Yang}, C.-C. 2023, \apj, 946, 60

\bibitem[{{Machida} {et~al.}(2010){Machida}, {Kokubo}, {Inutsuka}, \& {Matsumoto}}]{machida2010_gasacc}
{Machida}, M.~N., {Kokubo}, E., {Inutsuka}, S.-I., \& {Matsumoto}, T. 2010, \mnras, 405, 1227

\bibitem[{{Magg} {et~al.}(2022){Magg}, {Bergemann}, {Serenelli}, {Bautista}, {Plez}, {Heiter}, {Gerber}, {Ludwig}, {Basu}, {Ferguson}, {Gallego}, {Gamrath}, {Palmeri}, \& {Quinet}}]{magg2022_solarcomp}
{Magg}, E., {Bergemann}, M., {Serenelli}, A., {et~al.} 2022, \aap, 661, A140

\bibitem[{{Manara} {et~al.}(2023){Manara}, {Ansdell}, {Rosotti}, {Hughes}, {Armitage}, {Lodato}, \& {Williams}}]{manara2023_pp7discs}
{Manara}, C.~F., {Ansdell}, M., {Rosotti}, G.~P., {et~al.} 2023, in Astronomical Society of the Pacific Conference Series, Vol. 534, Protostars and Planets VII, ed. S.~{Inutsuka}, Y.~{Aikawa}, T.~{Muto}, K.~{Tomida}, \& M.~{Tamura}, 539

\bibitem[{{Manara} {et~al.}(2012){Manara}, {Robberto}, {Da Rio}, {Lodato}, {Hillenbrand}, {Stassun}, \& {Soderblom}}]{manara2012_accretion}
{Manara}, C.~F., {Robberto}, M., {Da Rio}, N., {et~al.} 2012, \apj, 755, 154

\bibitem[{{Moe} \& {Kratter}(2021)}]{morekratter2021_binariesSE}
{Moe}, M. \& {Kratter}, K.~M. 2021, \mnras, 507, 3593

\bibitem[{{Mori} {et~al.}(2019){Mori}, {Bai}, \& {Okuzumi}}]{mori2019_viscousheating}
{Mori}, S., {Bai}, X.-N., \& {Okuzumi}, S. 2019, \apj, 872, 98

\bibitem[{{Mulders} {et~al.}(2021){Mulders}, {Dr{\k{a}}{\.z}kowska}, {van der Marel}, {Ciesla}, \& {Pascucci}}]{mulders2021_MdwarfSE}
{Mulders}, G.~D., {Dr{\k{a}}{\.z}kowska}, J., {van der Marel}, N., {Ciesla}, F.~J., \& {Pascucci}, I. 2021, \apjl, 920, L1

\bibitem[{{Mulders} {et~al.}(2015){Mulders}, {Pascucci}, \& {Apai}}]{mulders2015_SEvMstar}
{Mulders}, G.~D., {Pascucci}, I., \& {Apai}, D. 2015, \apj, 798, 112

\bibitem[{{Mustill} {et~al.}(2015){Mustill}, {Davies}, \& {Johansen}}]{mustill2015_HJclearingSE}
{Mustill}, A.~J., {Davies}, M.~B., \& {Johansen}, A. 2015, \apj, 808, 14

\bibitem[{{Najita} \& {Bergin}(2018)}]{najitabergin2018_alpha}
{Najita}, J.~R. \& {Bergin}, E.~A. 2018, \apj, 864, 168

\bibitem[{{Nielsen} {et~al.}(2023){Nielsen}, {Gent}, {Bergemann}, {Eitner}, \& {Johansen}}]{nielsen2023}
{Nielsen}, J., {Gent}, M.~R., {Bergemann}, M., {Eitner}, P., \& {Johansen}, A. 2023, \aap, 678, A74

\bibitem[{{Nielsen} {et~al.}(2025){Nielsen}, {Johansen}, {Bali}, \& {Dorn}}]{nielsen2025}
{Nielsen}, J., {Johansen}, A., {Bali}, K., \& {Dorn}, C. 2025, \aap, 695, A184

\bibitem[{{Ohashi} {et~al.}(2023){Ohashi}, {Tobin}, {J{\o}rgensen}, {Takakuwa}, {Sheehan}, {Aikawa}, {Li}, {Looney}, {Williams}, {Aso}, {Sharma}, {Sai}, {Yamato}, {Lee}, {Tomida}, {Yen}, {Encalada}, {Flores}, {Gavino}, {Kido}, {Han}, {Lin}, {Narayanan}, {Phuong}, {Santamar{\'\i}a-Miranda}, {Thieme}, {van't Hoff}, {de Gregorio-Monsalvo}, {Koch}, {Kwon}, {Lai}, {Lee}, {Plunkett}, {Saigo}, {Hirano}, {Lam}, \& {Mori}}]{ohashi2023_edisk}
{Ohashi}, N., {Tobin}, J.~J., {J{\o}rgensen}, J.~K., {et~al.} 2023, \apj, 951, 8

\bibitem[{{Onyett} {et~al.}(2023){Onyett}, {Schiller}, {Makhatadze}, {Deng}, {Johansen}, \& {Bizzarro}}]{onyett2023_pebSS}
{Onyett}, I.~J., {Schiller}, M., {Makhatadze}, G.~V., {et~al.} 2023, \nat, 619, 539

\bibitem[{{Owen} {et~al.}(2012){Owen}, {Clarke}, \& {Ercolano}}]{owen2012_photoevap}
{Owen}, J.~E., {Clarke}, C.~J., \& {Ercolano}, B. 2012, \mnras, 422, 1880

\bibitem[{{Paardekooper} {et~al.}(2023){Paardekooper}, {Dong}, {Duffell}, {Fung}, {Masset}, {Ogilvie}, \& {Tanaka}}]{paardekooper2023_pp7migreview}
{Paardekooper}, S., {Dong}, R., {Duffell}, P., {et~al.} 2023, in Astronomical Society of the Pacific Conference Series, Vol. 534, Protostars and Planets VII, ed. S.~{Inutsuka}, Y.~{Aikawa}, T.~{Muto}, K.~{Tomida}, \& M.~{Tamura}, 685

\bibitem[{{Paardekooper} {et~al.}(2011){Paardekooper}, {Baruteau}, \& {Kley}}]{paardekooper2011_migration}
{Paardekooper}, S.~J., {Baruteau}, C., \& {Kley}, W. 2011, \mnras, 410, 293

\bibitem[{{Pan} {et~al.}(2025){Pan}, {Liu}, {Jiang}, {Xie}, {Zhu}, \& {Ribas}}]{pan2025_SEvMstar}
{Pan}, M., {Liu}, B., {Jiang}, L., {et~al.} 2025, \apj, 985, 7

\bibitem[{{Pan} {et~al.}(2024){Pan}, {Liu}, {Johansen}, {Ogihara}, {Wang}, {Ji}, {Wang}, {Feng}, \& {Ribas}}]{pan2024_collisions}
{Pan}, M., {Liu}, B., {Johansen}, A., {et~al.} 2024, \aap, 682, A89

\bibitem[{{Pascucci} {et~al.}(2016){Pascucci}, {Testi}, {Herczeg}, {Long}, {Manara}, {Hendler}, {Mulders}, {Krijt}, {Ciesla}, {Henning}, {Mohanty}, {Drabek-Maunder}, {Apai}, {Sz{\H{u}}cs}, {Sacco}, \& {Olofsson}}]{pascucci2016_dustmasses}
{Pascucci}, I., {Testi}, L., {Herczeg}, G.~J., {et~al.} 2016, \apj, 831, 125

\bibitem[{{Picogna} {et~al.}(2021){Picogna}, {Ercolano}, \& {Espaillat}}]{picogna2021_discPE}
{Picogna}, G., {Ercolano}, B., \& {Espaillat}, C.~C. 2021, \mnras, 508, 3611

\bibitem[{{Pinte} {et~al.}(2016){Pinte}, {Dent}, {M{\'e}nard}, {Hales}, {Hill}, {Cortes}, \& {de Gregorio-Monsalvo}}]{pinte2016_alpha}
{Pinte}, C., {Dent}, W.~R.~F., {M{\'e}nard}, F., {et~al.} 2016, \apj, 816, 25

\bibitem[{{Piso} \& {Youdin}(2014)}]{piso_youdin2014}
{Piso}, A.-M.~A. \& {Youdin}, A.~N. 2014, \apj, 786, 21

\bibitem[{{Reffert} {et~al.}(2015){Reffert}, {Bergmann}, {Quirrenbach}, {Trifonov}, \& {K{\"u}nstler}}]{reffert2015_GploccvMstar}
{Reffert}, S., {Bergmann}, C., {Quirrenbach}, A., {Trifonov}, T., \& {K{\"u}nstler}, A. 2015, \aap, 574, A116

\bibitem[{{Rosotti}(2023)}]{rosotti2023_alphaobs}
{Rosotti}, G.~P. 2023, \nar, 96, 101674

\bibitem[{{Shakura} \& {Sunyaev}(1973)}]{shakurasunyaev1973}
{Shakura}, N.~I. \& {Sunyaev}, R.~A. 1973, \aap, 24, 337

\bibitem[{{Shibata} \& {Helled}(2025)}]{shibatahelled2025_GPform}
{Shibata}, S. \& {Helled}, R. 2025, \aap, 700, A224

\bibitem[{{Stammler} {et~al.}(2023){Stammler}, {Lichtenberg}, {Dr{\k{a}}{\.z}kowska}, \& {Birnstiel}}]{stammler2023_fluxblock}
{Stammler}, S.~M., {Lichtenberg}, T., {Dr{\k{a}}{\.z}kowska}, J., \& {Birnstiel}, T. 2023, \aap, 670, L5

\bibitem[{{Tabone} {et~al.}(2025){Tabone}, {Rosotti}, {Trapman}, {Pinilla}, {Pascucci}, {Somigliana}, {Alexander}, {Vioque}, {Anania}, {Kuznetsova}, {Zhang}, {P{\'e}rez}, {Cieza}, {Carpenter}, {Deng}, {Agurto-Gangas}, {Ruiz-Rodriguez}, {Sierra}, {Kurtovic}, {Miley}, {Gonz{\'a}lez-Ruilova}, {TorresVillanueva}, {Hogerheijde}, {Schwarz}, {Toci}, {Testi}, \& {Lodato}}]{tabone2025_accmech}
{Tabone}, B., {Rosotti}, G.~P., {Trapman}, L., {et~al.} 2025, \apj, 989, 7

\bibitem[{{Trapman} {et~al.}(2020){Trapman}, {Rosotti}, {Bosman}, {Hogerheijde}, \& {van Dishoeck}}]{trapman2020_alpha}
{Trapman}, L., {Rosotti}, G., {Bosman}, A.~D., {Hogerheijde}, M.~R., \& {van Dishoeck}, E.~F. 2020, \aap, 640, A5

\bibitem[{{Van Clepper} {et~al.}(2025){Van Clepper}, {Price}, \& {Ciesla}}]{vanclepper2025_dustflowgap}
{Van Clepper}, E., {Price}, E.~M., \& {Ciesla}, F.~J. 2025, \apj, 980, 201

\bibitem[{{Van Zandt} {et~al.}(2025){Van Zandt}, {Petigura}, {Lubin}, {Weiss}, {Turtelboom}, {Fetherolf}, {Murphy}, {Crossfield}, {Gilbert}, {Mo{\v{c}}nik}, {Batalha}, {Dressing}, {Fulton}, {Howard}, {Huber}, {Isaacson}, {Kane}, {Robertson}, {Roy}, {Angelo}, {Behmard}, {Beard}, {Chontos}, {Dai}, {Giacalone}, {Hill}, {Holcomb}, {Howell}, {Mayo}, {Pidhorodetska}, {Polanski}, {Rogers}, {Rosenthal}, {Rubenzahl}, {Scarsdale}, {Tyler}, {Yee}, \& {Zink}}]{vanzandt2025_innerSE_outerGP}
{Van Zandt}, J., {Petigura}, E.~A., {Lubin}, J., {et~al.} 2025, \aj, 169, 235

\bibitem[{{Venturini} {et~al.}(2020){Venturini}, {Guilera}, {Haldemann}, {Ronco}, \& {Mordasini}}]{venturini2020_water}
{Venturini}, J., {Guilera}, O.~M., {Haldemann}, J., {Ronco}, M.~P., \& {Mordasini}, C. 2020, \aap, 643, L1

\bibitem[{{Villenave} {et~al.}(2022){Villenave}, {Stapelfeldt}, {Duch{\^e}ne}, {M{\'e}nard}, {Lambrechts}, {Sierra}, {Flores}, {Dent}, {Wolff}, {Ribas}, {Benisty}, {Cuello}, \& {Pinte}}]{villenave2022_alpha}
{Villenave}, M., {Stapelfeldt}, K.~R., {Duch{\^e}ne}, G., {et~al.} 2022, \apj, 930, 11

\bibitem[{{Zink} {et~al.}(2023){Zink}, {Hardegree-Ullman}, {Christiansen}, {Petigura}, {Boley}, {Bhure}, {Rice}, {Yee}, {Isaacson}, {Fernandes}, {Howard}, {Blunt}, {Lubin}, {Chontos}, {Pidhorodetska}, \& {MacDougall}}]{zink2023_occ_vs_amplitude}
{Zink}, J.~K., {Hardegree-Ullman}, K.~K., {Christiansen}, J.~L., {et~al.} 2023, \aj, 165, 262

\end{thebibliography}
\begin{appendix}
\onecolumn
\section{Different photoevaporation models}
\label{app:photo_evap}
\begin{figure*}[h!]
    \centering
    \includegraphics[width=\linewidth]{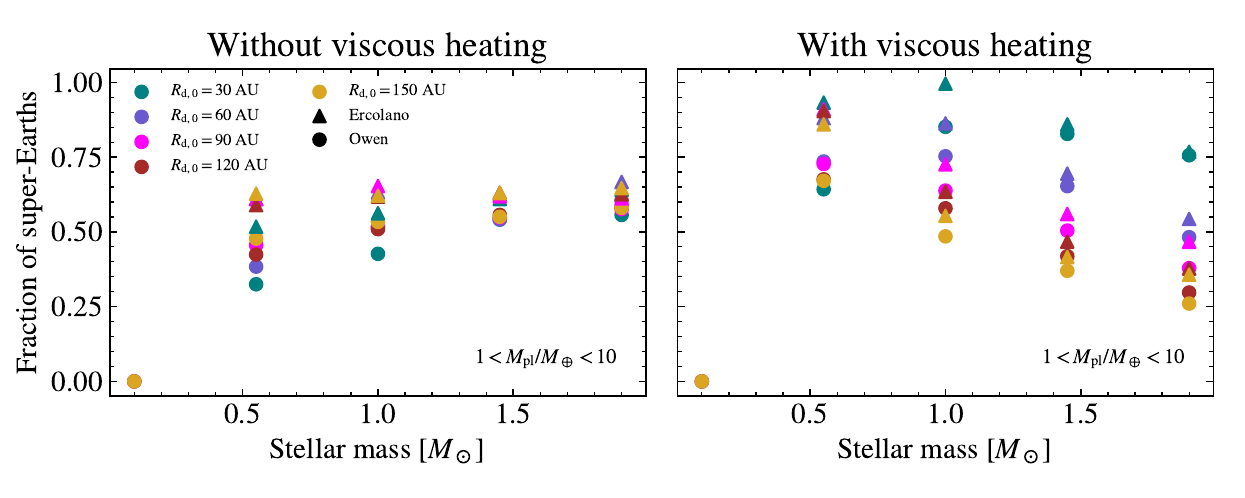}
    \caption{Close in super-Earth fraction as a function of stellar mass for different disc sizes. The circles show the results using our nominal photoevaporation model (same results as in fig. \ref{fig:scaling_disc_SE_close_frac}), while the triangles show the results using the photoevaporation model from \citet{ercolano2023_discPE}. We find no significant differences between the two models.}
    \label{fig:SE_close_frac_ercolano_compare}
\end{figure*}

In fig. \ref{fig:SE_close_frac_ercolano_compare}, we show the close-in super-Earth fraction using for both our nominal photoevaporation model from \citet{owen2012_photoevap} and the photoevaporation from \citet{ercolano2023_discPE}. The photoevaporation model from \citet{ercolano2023_discPE} results in shorter disc lifetimes due to the overall higher photoevaporation rates. However, as the surface density of the disc is low at later stages of its lifetime, extending the lifetime beyond a few megayears, as our nominal model does, does not affect the planet population. 
\twocolumn
\section{Migration coefficient as a function time and disc size}
\label{app:mig_coeff}
\begin{figure}[h!]
    \centering
    \includegraphics[width=\linewidth]{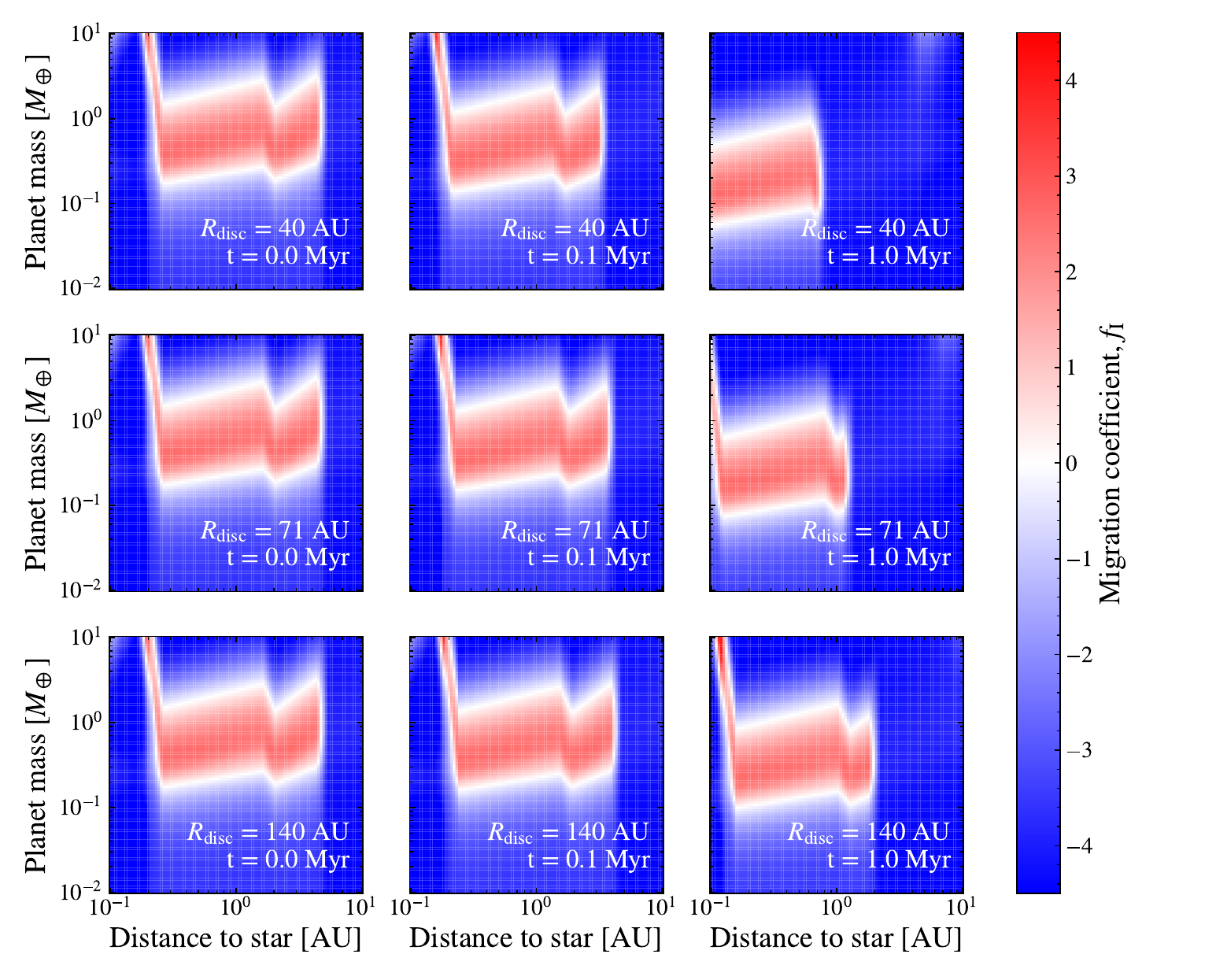}
    \caption{Same as Fig. \ref{fig:mig_coeff} but at different times and for different disc sizes.}
    \label{fig:mig_coeff_time_size_solar}
\end{figure}

\begin{figure}[h!]
    \centering
    \includegraphics[width=\linewidth]{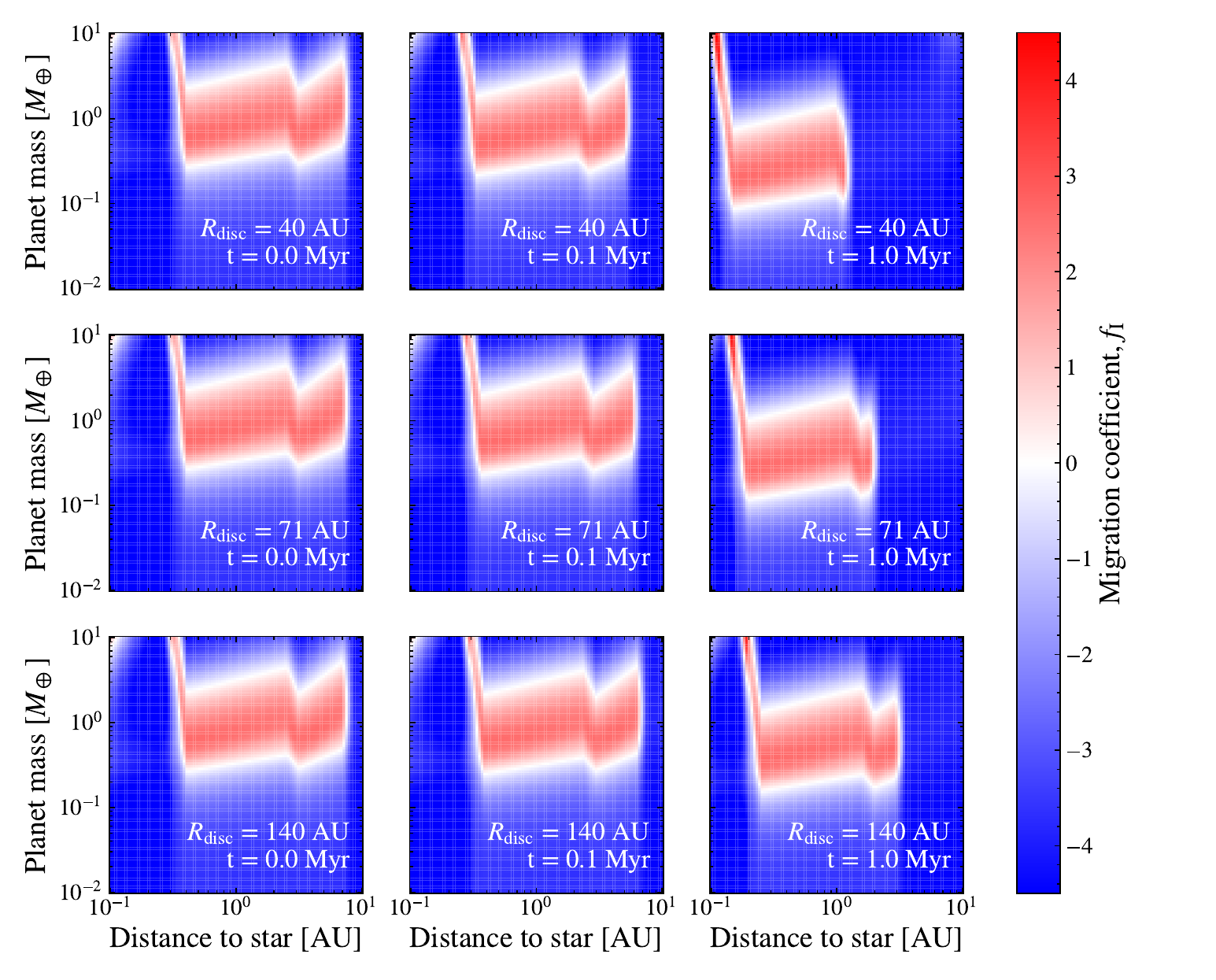}
    \caption{Same as Fig. \ref{fig:mig_coeff_time_size_solar} but for a star with a mass of 1.5 $M_\odot$.}
    \label{fig:mig_coeff_time_size_massive}
\end{figure}

\begin{figure}[h!]
    \centering
    \includegraphics[width=\linewidth]{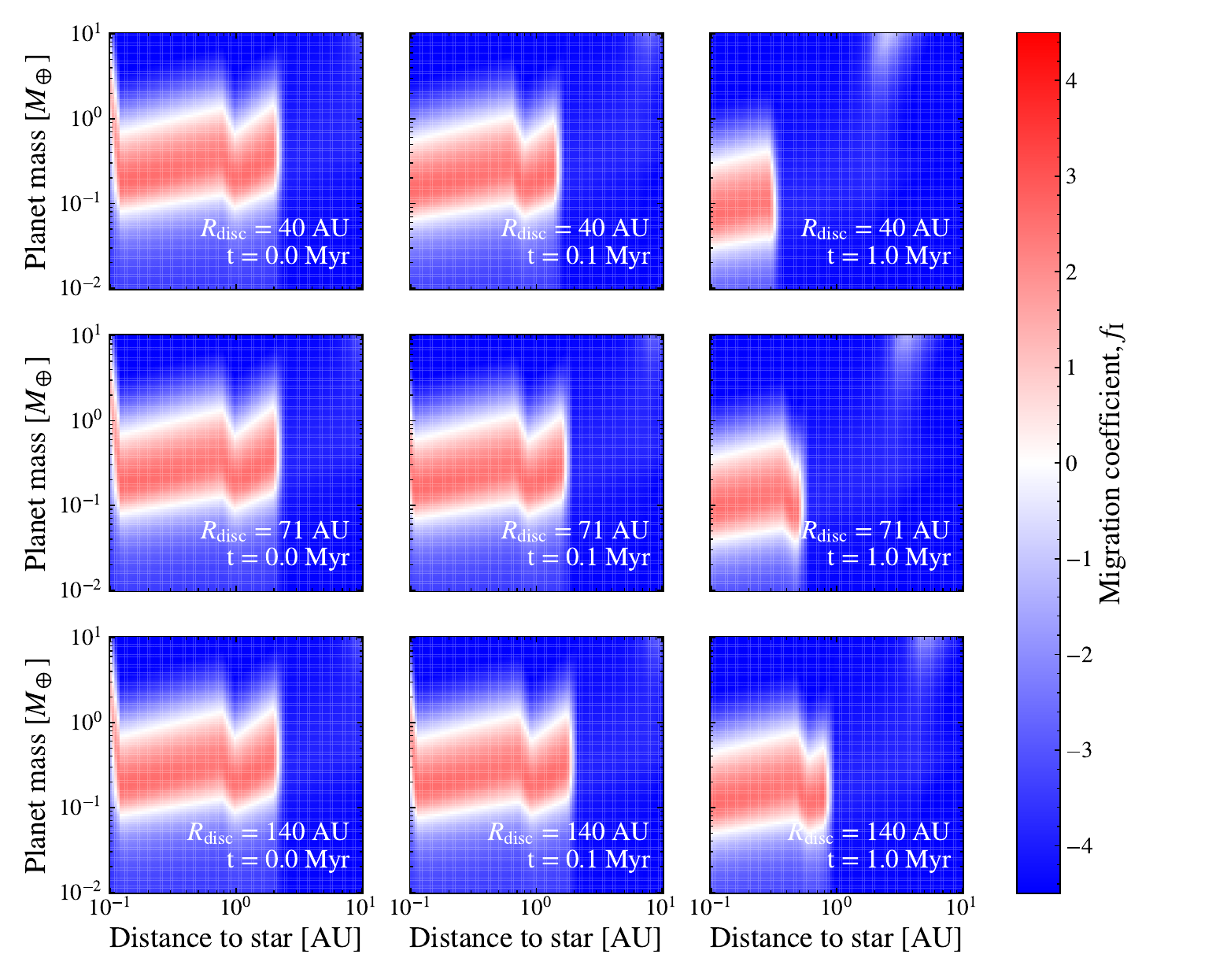}
    \caption{Same as Fig. \ref{fig:mig_coeff_time_size_solar} but for a star with a mass of 0.5 $M_\odot$.}
    \label{fig:mig_coeff_time_size_small}
\end{figure}

Figs. \ref{fig:mig_coeff_time_size_solar}-\ref{fig:mig_coeff_time_size_small} show the migration coefficient as a function of distance to the star and planet mass but for different stellar masses. There are no significant differences for different stellar masses with the exception of $r_{\rm tran}$ extending slightly farther out for larger stellar masses. As larger discs evolve slower than smaller disc, the region where outward migration is prevalent moves inwards faster for smaller discs, which results in more efficient inward migration around smaller discs. 
\onecolumn
\section{Additional simulations}
\label{app:additional_figs}
\begin{figure*}[h!]
    \centering
    \includegraphics[width=\linewidth]{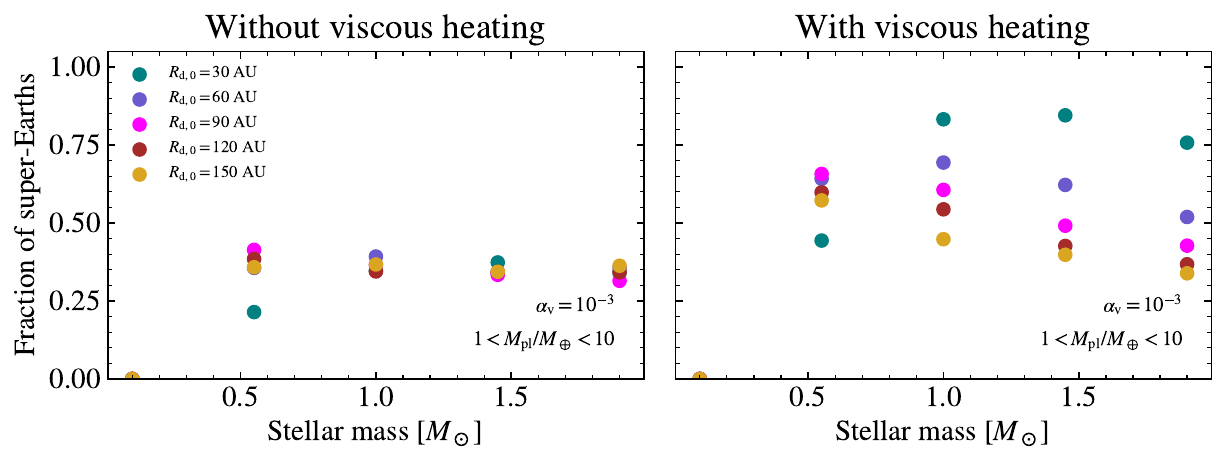}
    \caption{Same as Fig. \ref{fig:scaling_disc_SE_close_frac} but with $\alphav = 10^{-3}$ and $\dot{M}_{\rm g,0} = 10^{-8}\,M_\odot/{\rm yr}$. We find similar results in both the irradiated case (left) and the viscously heated case (right) compared to our nominal model. In the irradiated case, the fraction of super-Earths saturates at a lower stellar mass due to efficient inward migration. In the viscously heated case, we find that the fraction of close-in super-Earth still decreases with increasing stellar mass for large discs but the reduction is slightly lower compared to our nominal model.}
    \label{fig:scaling_disc_lowalpha}
\end{figure*}

\begin{figure*}[h!]
    \centering
    \includegraphics[width=\linewidth]{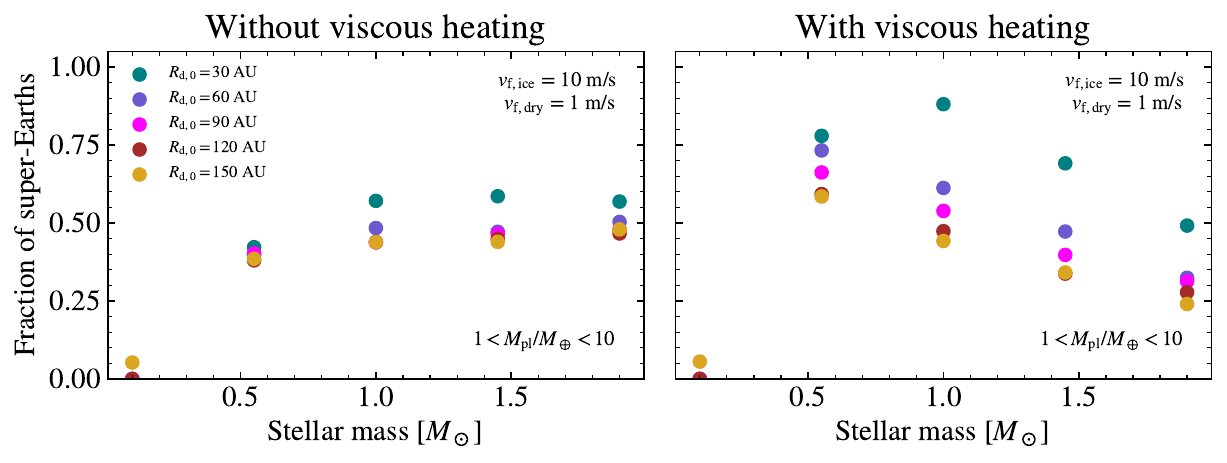}
    \caption{Same as Fig. \ref{fig:scaling_disc_SE_close_frac} but with pebble fragmentation velocities set to 1 m/s for dry pebble inside the water iceline and 10 m/s for icy pebbles outside the water iceline. The reduction of super-Earths for higher stellar masses is slightly stronger in the viscous case, as the dry region of the disc -- where pebbles are small and planet formation is inefficient -- extends farther away from the star for higher stellar masses. In the irradiated case, the results are relatively unchanged.}
    \label{fig:scaling_disc_varied_vf}
\end{figure*}

In Figs. \ref{fig:scaling_disc_lowalpha} and \ref{fig:scaling_disc_varied_vf}, we show the close-in super-Earth fraction for a different value of $\alphav$ and different fragmentation velocities, respectively. We find minimal differences between these results and our nominal model.
\end{appendix}
\end{document}